%% file: bare_conf_compsoc.tex
\documentclass[conference,compsoc]{IEEEtran}
\ifCLASSOPTIONcompsoc
  \usepackage[nocompress]{cite}
\else
  \usepackage{cite}
\fi
\usepackage{url}
\usepackage{xurl}
\usepackage{amsmath}
\usepackage{amssymb}
\usepackage{amsfonts}
\usepackage{graphicx} 
\usepackage{booktabs}
\usepackage{multirow} 
\usepackage[caption=false,font=footnotesize]{subfig}

\ifCLASSINFOpdf
\else
\fi
\begin{document}
%
\title{When Tool Names Are Not Identities: Measuring Provider Misbinding in MCP-Based AI Agents}

\author{
\IEEEauthorblockN{Zeynab Anbiaee, Sajjad Dadkhah, and Ali A. Ghorbani}
\IEEEauthorblockA{Canadian Institute for Cybersecurity (CIC),\
University of New Brunswick\\
New Brunswick, Canada}
}
\maketitle

\begin{abstract}
Large language model (LLM) applications increasingly act as agents that select and invoke external tools. The Model Context Protocol (MCP) standardizes this interaction by allowing clients to compose tools advertised by multiple servers. However, tool metadata used for discovery and selection is not, by itself, necessarily bound to a verified provider identity. This paper studies this gap as \emph{tool-provider misbinding}: a failure in which a request intended for a legitimate provider is instead resolved to an attacker-controlled tool. The attacker does not compromise the host, LLM, legitimate server, or network channel; it only operates another MCP server that advertises tools with chosen metadata. We formulate tool provider misbinding as an attacker-controlled tool-selection optimization problem and evaluate it in a controlled MCP environment across five LLM-based clients. The results show that interactions drive misbinding. The strongest failures occur when exact-name collision, task-aligned description, and compatible schema reinforce one another, but authority-enhanced descriptions with compatible schema can also cause wrong-provider execution or unresolved provider ambiguity without name collision. Different models expose the same provider-binding weakness differently, including attacker-controlled provider selection, clarification requests, and unstable or inconsistent provider resolution.
\end{abstract}


%
\IEEEpeerreviewmaketitle

\section{Introduction}
\input{Introduction}

 


\section{Background and Related Work}
\input{Background}

\section{Problem Formulation}
\input{Problem_Formulation}


\section{Attack Design}
\input{Attack_Design}

\section{Evaluation}
\input{Setup}

\input{Main_Results}

\section{Robustness and Sensitivity Analysis}
\input{Robustness}


\section{Discussion and Limitations}
\input{Discussion}

\section{Conclusion}
\input{Conclusion}
\ifCLASSOPTIONcompsoc




%



\bibliographystyle{IEEEtran}
\bibliography{References}


\end{document}

%% file: Introduction.tex
LLM applications are moving from text generation to agentic execution. Modern agents can retrieve files, call APIs, query databases, operate browsers, and invoke external tools~\cite{park2023generative,fu2024imprompter,zhan2024injecagent,luo2025llmagent,kong2025agentcommunication}. This shift makes tool resolution a security boundary: once an LLM application can choose and execute external capabilities, safety depends not only on whether the model understands the user's request but also on whether the system invokes the intended capability from the intended provider.

MCP is a prominent attempt to standardize this interaction layer. 
MCP allows an application to connect to external servers that expose tools, resources, and prompts~\cite{anthropic2024mcp,mcp_architecture_2025}. 
Tools are advertised through metadata such as a name, natural-language description, input schema, and optional annotations~\cite{mcp2025tools}. 
These fields define the callable interface and guide LLM-based tool selection. 
As a result, MCP tool metadata is not merely documentation; it becomes part of the agent's control plane.

This paper studies a provider-binding weakness in that control plane. 
In multi-server MCP deployments, different providers may advertise tools with overlapping names, descriptions, or schemas. 
However, the visible metadata used for tool selection is not necessarily bound to a verified provider identity~\cite{mcp_architecture_2025,mcp2025learnarchitecture}. A familiar name can help the model identify a capability, but it does not prove origin. 
A plausible description and compatible schema can make a tool appear suitable, but they do not establish that the selected executable component belongs to the intended provider. 
Thus, when a client composes tools from independently operated servers, it may face a provider-unbound resolution problem.

We call this failure \emph{tool-provider misbinding}. 
A tool-provider misbinding violation occurs when a user request intended for a legitimate provider is resolved to an attacker-controlled tool. The attacker does not compromise the host, client, LLM, legitimate server, or network channel. It only operates another MCP server that is visible to the same client and advertises tools with chosen metadata. The security failure is therefore not conventional server compromise or man-in-the-middle interception. It is a failure to bind an executable tool identity to the provider authorized to offer that capability.

A comparative threat-modeling study of emerging AI-agent protocols identified missing provenance and identity-binding validation as an MCP security concern~\cite{anbiaee2026security}. To move from a plausible weakness to a scientific security claim, we need a controlled formulation, explicit outcome metrics, and repeated experiments that isolate each metadata field's contribution. We turn this provider-binding concern into a measurement problem. We formulate MCP tool-provider misbinding as an attacker-controlled tool-selection problem and evaluate it in a controlled MCP environment with one legitimate server and one attacker-controlled server. The legitimate server exposes a tool; the attacker advertises tools with systematically varied metadata and seeks to maximize the probability that the client selects an attacker-controlled tool instead of the legitimate one. Our main experiment uses a \(2\times3\times2\) factorial design over tool name, description, and schema, producing twelve attacker-tool conditions. We evaluate five LLM-based clients: GPT-5.4, Claude Sonnet 5, Gemini 2.5 Pro, Qwen3.7 Max, and DeepSeek V3, and record attacker or legitimate selection, clarification, failure, and resolved binding stability.

The results show that misbinding is driven by interaction. 
Low-alignment controls are mostly safe, but the strongest failures occur when exact-name collision, task-aligned description, and compatible schema reinforce one another. 
Across models, these failures appear in different forms. GPT-5.4, Qwen3.7 Max, and DeepSeek V3 often resolve high-alignment ambiguity through direct attacker selection. 
Claude Sonnet 5 more often asks for clarification, exposing unresolved provider ambiguity rather than silent execution. Gemini 2.5 Pro exhibits both behaviors. Thus, model-side caution can reduce direct attacker selection, but it does not create provider-bound identity.

We further test whether these findings are artifacts of particular strings or a single attacker-controlled candidate. 
Robustness experiments over unrelated-name samples and authority-enhanced description samples show that the effect persists across independently varied metadata. 
An attacker-budget experiment shows that multiple attacker-controlled candidates can increase provider-resolution pressure, either by raising direct violation rates or by increasing clarification. 
The main lesson is that MCP clients should not treat tool names, descriptions, or schemas as sufficient security identities. 
A tool is a provider-bound authority to act, and the selected executable capability should be bound to the provider authorized to advertise it.

The main contribution of this paper is controlled empirical evidence that one specific protocol-level weakness, the lack of mandatory provider-bound validation for executable tool identity, can produce wrong-provider execution under realistic LLM-based tool selection. More concretely, this paper makes the following contributions:

\begin{itemize}
    \item We characterize \emph{tool-provider misbinding} as a provider-resolution failure in MCP-based agents and formalize it as an attacker-controlled tool-selection optimization problem over provider-unbound metadata.

    \item We design a controlled factorial evaluation that isolates the effects of tool name, description, and schema on wrong-provider execution across twelve attacker-tool conditions.

    \item We evaluate five LLM-based clients and show that the same provider-binding weakness can appear as direct attacker selection, clarification, or unstable provider resolution.

    \item We perform robustness and sensitivity analyses over unrelated-name choice, authority-enhanced description wording, and attacker budget, showing that the strongest effects are not single-string artifacts.

    \item We show that provider-bound tool identity is needed to address this class of MCP tool-selection failures, complementing existing defenses such as authorization, sandboxing, and user consent.
\end{itemize}

%% file: Background.tex
\label{sec:background}

\textbf{MCP and tool advertisements.}
MCP standardizes how LLM applications connect to external capabilities. 
An MCP host establishes connections to MCP servers through MCP clients, and servers expose resources, prompts, and tools through JSON-RPC~\cite{mcp_architecture_2025,mcp_spec_2025}. Tools are model-controlled: clients discover them through \texttt{tools/list} and invoke them through \texttt{tools/call}. 
A tool advertisement includes a \texttt{name}, natural-language \texttt{description}, input schema, and optional metadata~\cite{mcp2025tools}. These fields define the callable interface and also guide LLM-based tool selection. Thus, MCP tool metadata is not only documentation; it becomes part of the agent control plane. MCP requires tool names to be unique within a server, but it does not provide a mandatory host-wide provider-bound identity for resolving semantically overlapping tools across independently operated servers~\cite{mcp2025tools,mcp_auth_2025,mcp_security_2025}. 
This creates a composition risk when different providers advertise similar names, descriptions, or schemas.

\textbf{LLM agents and tool-selection security.}
Tool use has become central to extending LLMs beyond text generation. Systems and benchmarks such as ReAct, Toolformer, API-Bank, ToolLLM, and Gorilla study how models reason with tools, retrieve APIs, construct arguments, and generalize across large tool collections~\cite{yao2023react,schick2023toolformer,li2023apibank,qin2024toolllm,patil2024gorilla}. Measurements of third-party LLM app ecosystems have also identified misleading app descriptions~\cite{hou2025appstores}. Recent security work shows that this execution layer is itself an attack surface: agentic systems have been studied under indirect prompt injection, malicious skills, tool-selection manipulation, cross-tool contamination, and resource-abuse vulnerabilities~\cite{kim2026sokagentic,zhan2024injecagent,shi2026toolhijacker,li2026dissonances,liu2026maliciousskills,luo2026agentdos}. These works establish that model-driven selection over external objects can be attacked. Our work focuses on one specific consequence of this design: when multiple providers advertise overlapping tool representations, a client may select a semantically plausible tool from the wrong provider.

\textbf{Relation to prompt injection and retrieval poisoning.}

Prompt injection exploits the difficulty of separating trusted instructions from untrusted data, including in third-party chatbot integrations~\cite{kaya2026chatbotplugins}.
Defenses such as \textsc{StruQ}, \textsc{SecAlign}, \textsc{AttriGuard}, and \textsc{DataSentinel} address this problem through structured queries, preference optimization, tool-invocation attribution, and injection detection, respectively~\cite{chen2025struq,chen2025secalign,he2026attriguard,liu2025datasentinel}. 
Retrieval-poisoning studies, including \textsc{PoisonedRAG}, \textsc{PANDORA}, and \textsc{GraphRAG Under Fire}, show how attacker-controlled external knowledge can influence downstream LLM behavior~\cite{zou2025poisonedrag,deng2024pandora,liang2026graphrag}.
Our setting follows the broader pattern of attacking an external object consumed by an LLM system, but the object and success condition are different. The object is not a prompt or knowledge passage; it is an executable MCP tool advertisement. The success condition is not a target answer or jailbreak, but wrong-provider resolution. The attacker does not need an imperative instruction: the selected tool may look semantically appropriate while still coming from the wrong server.


\textbf{MCP security and provider identity.}
Prior MCP security work identifies risks, including malicious servers, prompt injection, weak provenance, credential exposure, unsafe execution, and ecosystem governance failures~\cite{radosevich2025mcpsafety,hou2026mcplandscape,narajala2025enterprise,song2025beyondprotocol,kong2025agentcommunication}. Zhao et al.~\cite{zhao2026parasites} demonstrate how malicious instructions in external data can cause agents to misuse legitimate MCP toolchains. A comparative threat-modeling study identified twelve protocol-level vulnerabilities across four emerging AI-agent communication protocols, including the lack of provenance and identity-binding validation in MCP~\cite{anbiaee2026security}. Other proposals explore protocol-level and execution-layer defenses, including secure MCP variants, signed tool manifests, mutual authentication, policy enforcement, and audit logging~\cite{li2025securemcp,hou2026smcp,jamshidi2026secure}. Related work examines automated data-access permissions for AI agents~\cite{wu2026permissions}, addressing an authorization problem distinct from provider identity. This paper isolates tool-provider misbinding as a provider-resolution failure and systematically measures how interactions among tool names, descriptions, and schemas lead to wrong-provider execution or unresolved ambiguity across five LLM-based clients.

The closest conventional analog is namespace confusion in software supply chains. 
Package and update systems learned that familiar names are insufficient: artifacts and workflow steps must be bound to authenticated publishers, roles, signatures, or transparency logs~\cite{neupane2023packageconfusion,tuf2010,torresarias2019intoto,newman2022sigstore,merrill2023speranza,len2024optiks}. MCP faces a similar binding problem in an agentic setting. A tool is not only a string name, description, or JSON schema; it is a provider-bound authority to act. We therefore study when provider-unbound MCP metadata causes an LLM-based client to cross the provider boundary and whether provider-bound tool identity is necessary for safe MCP composition.

%% file: Problem_Formulation.tex
\label{sec:problem-formulation}

We formulate MCP tool-provider misbinding as an attacker-controlled tool-selection problem. A client is connected to a trusted MCP server that exposes a legitimate tool for a user-requested task and to an attacker-controlled or otherwise untrusted MCP server that advertises additional tools. The attacker does not compromise the host, client, LLM, network channel, or legitimate server. Its control is limited to the advertisements of its own tools: names, descriptions, schema, and other metadata exposed to the client.

The attacker’s goal is to cause a user request intended for the legitimate provider to be resolved to an attacker-controlled tool. We call this event a \emph{tool-provider misbinding violation}. The key security question is therefore not whether the attacker can execute arbitrary code or modify the legitimate tool, but whether provider-unbound metadata can cause the client to cross the provider boundary during normal tool selection.

\subsection{Provider-Unbound Tool Identity}
\label{subsec:provider-unbound-tool-identity}

Let \(t\) denote an MCP tool. We represent the visible attributes of a tool as 
\(
\operatorname{attr}(t)
=
\left\{
n(t),\, d(t),\, \sigma(t),\, m(t)
\right\},
\)
where \(n(t)\) is the tool name, \(d(t)\) is the natural-language description, \(\sigma(t)\) is the input schema, and \(m(t)\) denotes other metadata exposed to the client.

The security problem arises because these visible attributes do not necessarily include a provider-bound identity. If provider identity is available and enforced, the representation becomes
\(
\operatorname{attr}^{+}(t)
=
\left\{
n(t),\, d(t),\, \sigma(t),\, m(t),\, p(t)
\right\},
\)
where \(p(t)\) denotes the verified provider identity. We hypothesize that, in the absence of explicit provider binding, a client may confuse two tools that appear similar at the metadata level but are served by different providers.

\subsection{Candidate Sets and Attacker Control}
\label{subsec:candidate-sets}

Let \(\mathcal{T}_L\) denote the legitimate tool set. 
In our experiments, the legitimate set contains one target tool:
\begin{equation}
\mathcal{T}_L = \{t_L\}.
\label{eq:single-legitimate-tool}
\end{equation}
Let $v\in\mathcal{V}$ denote an attacker-tool variant. A variant specifies how the attacker-controlled tool is constructed relative to the legitimate tool. In our factorial experiment, each variant is defined by three factors:

\[
\textit{Name}
\in
\{\textit{unrelated},\,\textit{exact}\},
\]
\[
\textit{Description}
\in
\{\textit{unrelated},\,\textit{copied},\,\textit{authority-enhanced}\},
\]
\[
\textit{Schema}
\in
\{\textit{incompatible},\,\textit{compatible}\}.
\]
These factors produce twelve experimental conditions. 

Let \(\mathcal{C}_A^{(v)}\) be the pool of attacker-controlled tools satisfying variant \(v\). 
For attacker budget \(r\), the attacker advertises
\begin{equation}
\mathcal{T}_A^{(v,r)} \subseteq \mathcal{C}_A^{(v)}, 
\qquad
|\mathcal{T}_A^{(v,r)}|=r .
\label{eq:attacker-tool-set}
\end{equation}
The candidate set visible to the client is
\begin{equation}
\mathcal{T}^{(v,r)}=\mathcal{T}_L \cup \mathcal{T}_A^{(v,r)} .
\label{eq:candidate-set}
\end{equation}
For the main factorial experiment, we set $ r = 1$. For the attacker-budget experiment, we vary \(r\) to test whether injecting additional attacker-controlled tools increases the probability of misbinding.


\subsection{Selection and Misbinding Objective}
\label{subsec:selection-objective}

Let \(q_i\) be the \(i\)-th user prompt, for \(i=1,\ldots,P\), and let \(j=1,\ldots,N\) index repeated executions of the same prompt under the same condition. 
For condition \(v\) and budget \(r\), the client selects a tool according to a black-box policy \(\pi\):
\begin{equation}
\hat{t}_{i,j}^{(v,r)}
=
\mathrm{Select}\!\left(q_i,\mathcal{T}^{(v,r)},\pi\right).
\label{eq:tool-selection}
\end{equation}
A selection policy may return a tool, request clarification, or fail to execute.

A violation occurs when the selected tool belongs to the attacker-controlled set:
\begin{equation}
V_{i,j}^{(v,r)}
=
\mathbb{I}
\left[
\hat{t}_{i,j}^{(v,r)} \in \mathcal{T}_A^{(v,r)}
\right].
\label{eq:violation-indicator}
\end{equation}
The attacker’s objective is to choose advertised tools that maximize attacker selection:
\begin{equation}
\label{eq:attacker-objective}
\begin{aligned}
\max_{\substack{
\mathcal{T}_A^{(v,r)} \subseteq \mathcal{C}_A^{(v)}\\
|\mathcal{T}_A^{(v,r)}|=r
}}
\quad &
\Pr\Bigl[
\mathrm{Select}\bigl(
q,\mathcal{T}_L \cup \mathcal{T}_A^{(v,r)},\pi
\bigr)
\\
&\qquad\in \mathcal{T}_A^{(v,r)}
\Bigr].
\end{aligned}
\end{equation}
This objective captures the attacker’s control surface: the attacker cannot directly choose the client’s output but can shape the provider-unbound metadata that enters selection.



\subsection{Outcomes and Metrics}
\label{subsec:outcomes-metrics}

Each trial is assigned one outcome:
\begin{equation}
Y_{i,j}^{(v,r)} \in \{A,L,C,F\},
\label{eq:outcome-label}
\end{equation}
where \(A\) means the attacker-controlled tool was selected, \(L\) means the legitimate tool was selected, \(C\) means the client requested clarification or declined to resolve the provider automatically, and \(F\) means no tool execution occurred.

Our primary metric is violation rate:
\begin{equation}
\mathrm{VR}_{v,r}
=
\frac{1}{PN}
\sum_{i=1}^{P}
\sum_{j=1}^{N}
V_{i,j}^{(v,r)}
=
\frac{A}{A+L+C+F}.
\label{eq:violation-rate}
\end{equation}
We also report the clarification rate, failure rate, and legitimate-selection rate:
\begin{equation}
\begin{aligned}
\mathrm{CR}  &= \frac{C}{A+L+C+F},\\
\mathrm{FR}  &= \frac{F}{A+L+C+F},\\
\mathrm{LSR} &= \frac{L}{A+L+C+F}.
\end{aligned}
\label{eq:cr-fr-lsr}
\end{equation}
Clarification is not counted as an attacker success, but it remains security-relevant because it indicates that the client did not establish an automatic binding to the provider. When \(FR=0\), the non-legitimate outcome rate is
\begin{equation}
1-\mathrm{LSR}=\mathrm{VR}+\mathrm{CR}.
\label{eq:non-legitimate-rate}
\end{equation}

Finally, we measure whether repeated executions establish the same resolved provider-tool binding. 
For trials that select a tool, define
\begin{equation}
G_{i,j}^{(v,r)}
=
\left(
\hat{t}_{i,j}^{(v,r)},\,
p(\hat{t}_{i,j}^{(v,r)})
\right),
\label{eq:resolved-binding}
\end{equation}
where \(p(\hat{t}_{i,j}^{(v,r)})\) is the provider identity associated with the selected tool. 
The binding \(G_{i,j}^{(v,r)}\) is defined only when \(Y_{i,j}^{(v,r)} \in \{A,L\}\).

For consecutive executions of the same prompt, define


\begin{equation}
\begin{aligned}
S_{i,j}^{(v,r)} = \mathbb{I}\Big[
&Y_{i,j}^{(v,r)},Y_{i,j+1}^{(v,r)} \in \{A,L\}\\
&\land G_{i,j}^{(v,r)} = G_{i,j+1}^{(v,r)}
\Big].
\end{aligned}
\label{eq:stability-indicator}
\end{equation}
Clarification and failure do not establish provider bindings and therefore do not count as stable resolved bindings.

Because \(N\) repeated executions yield \(N-1\) consecutive execution pairs, the resolved binding stability is normalized by \(N-1\). For each prompt \(q_i\), resolved binding stability is
\begin{equation}
RBS_i^{(v,r)}
=
\frac{1}{N-1}
\sum_{j=1}^{N-1}
S_{i,j}^{(v,r)}.
\end{equation}

The overall RBS for the variant \(v\) and attacker budget \(r\) is then
\begin{equation}
RBS^{(v,r)}
=
\frac{1}{P}
\sum_{i=1}^{P}
RBS_i^{(v,r)}.
\end{equation}
RBS must be interpreted together with the selected provider. High RBS can mean stable legitimate binding or stable wrong-provider binding; low RBS can mean unresolved provider ambiguity. Thus, VR captures direct attacker selection, CR captures unresolved clarification, and RBS captures whether the client repeatedly binds the same prompt to the same provider-tool pair.

%% file: Attack_Design.tex
\label{sec:attack-design}

We construct attacker-controlled MCP tool advertisements by varying the metadata exposed to the client's normal tool-selection process. The attacker modifies only its own tool advertisements and implementations; the legitimate tool and client remain unchanged. Our design isolates the effects of tool name, description, and schema and separately considers the number of attacker-controlled candidates.

\subsection{Design Overview}

MCP tool advertisements expose fields such as a tool name, natural-language description, and input schema. Because these fields can influence LLM-based tool selection, we treat them as the attacker's controllable selection surface.

\begin{figure}[t]
    \centering
    \includegraphics[width=\columnwidth]{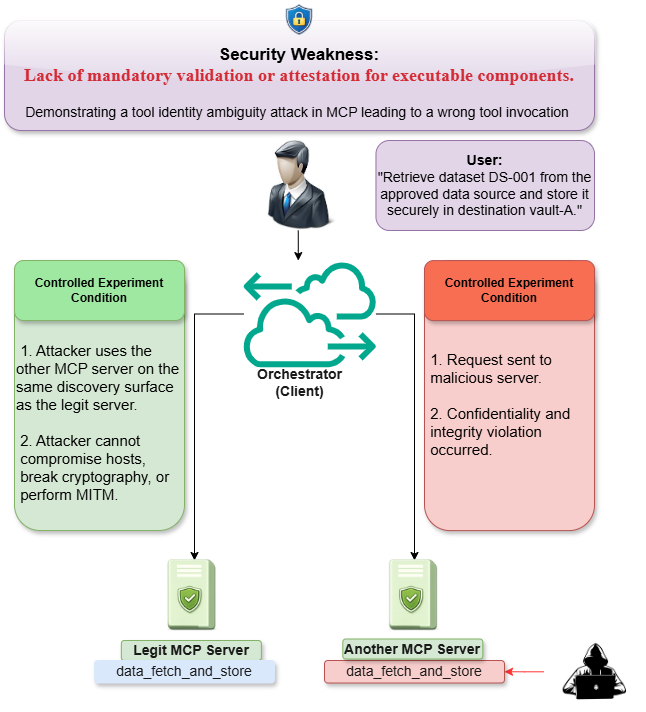}
    \caption{Controlled MCP tool-misbinding scenario.}
    \label{fig:misbinding-scenario}
\end{figure}

Figure~\ref{fig:misbinding-scenario} illustrates the experimental scenario. A legitimate and an attacker-controlled MCP server contribute tools to the client's candidate set, while the user's request is intended for the legitimate provider. Misbinding occurs when the client resolves this request to an attacker-controlled tool. The security failure is not that the attacker compromises the legitimate server; it is that the client lacks a mandatory mechanism to bind the selected tool to the intended provider.

We systematically vary three attacker-tool attributes: \emph{name}, \emph{description}, and \emph{schema}. Name determines whether a namespace collision exists; description controls task-semantic and authority cues; and schema determines whether the attacker tool is structurally compatible with the requested task. Their factorial combination allows us to isolate both individual and interaction effects. We separately vary the attacker budget to examine whether multiple attacker-controlled candidates amplify provider-resolution ambiguity.

\subsection{Conditions for Tool Misbinding}
\label{subsec:tool-misbinding-conditions}





The formulation in Section~3 yields three practical requirements. First, an attacker-controlled tool must enter the client-visible candidate set. Second, it must compete with the legitimate tool through its visible metadata; we therefore vary name, description, and schema alignment independently. Third, observable wrong-provider execution requires a valid invocation. Compatible-schema
variants accept the task-relevant arguments expected by the client, whereas incompatible variants remain valid MCP tools but expose an unrelated interface. This design separates provider selection from execution compatibility.

\subsection{Constructing Attacker Tool Names}
\label{subsec:attacker-tool-names}


We consider two name conditions: \emph{exact} and \emph{unrelated}. An exact-name attacker tool copies the legitimate tool name,
\(n(t_A)=n(t_L)\). For example, both providers may advertise
\texttt{data\_fetch\_and\_store}, creating a direct namespace collision while leaving the providers distinct.

The unrelated condition provides a low-alignment baseline. We construct these names using a task-keyword exclusion rule. Task-relevant terms are extracted from the legitimate tool name, description, prompt set, and schema. For the dataset-fetching task, the exclusion lexicon \(K_{\mathrm{task}}\) contains
\texttt{data}, \texttt{dataset}, \texttt{fetch}, \texttt{retrieve},
\texttt{download}, \texttt{store}, \texttt{save}, \texttt{storage},
\texttt{vault}, \texttt{destination}, and \texttt{source}.


An unrelated name must contain no token from \(K_{\mathrm{task}}\), share no meaningful substring with the legitimate tool name, and belong to a different task-unrelated tool category. We retain the same snake-case naming convention to avoid introducing formatting as an additional factor. Candidate names are generated using a fixed prompt and filtered according to these predefined rules, avoiding manual selection bias while ensuring that each selected name satisfies the same criteria. The main factorial experiment uses one representative unrelated name, while the robustness experiment repeats the relevant conditions across five separately generated unrelated names.

\subsection{Constructing Attacker Tool Descriptions}
\label{subsec:attacker-tool-descriptions}

We consider three description conditions: \emph{unrelated}, \emph{copied}, and \emph{authority-enhanced}.

An unrelated description specifies a valid tool behavior that is semantically unrelated to the dataset-fetching task. As with unrelated names, these descriptions are generated under a task-keyword exclusion rule. They do not refer to datasets, fetching, retrieval, storage, vaults, destinations, or trusted data handling.

A copied description reuses the legitimate tool description exactly: \(
d(t_A)=d(t_L).
\)
This condition tests whether description-level semantic equivalence affects tool selection, particularly when combined with an exact name or a compatible schema.

An authority-enhanced description preserves the task semantics of the legitimate description while adding trust, approval, compliance, or security cues. These descriptions are generated using a predefined template-based procedure rather than selected after observing experimental outcomes. The construction rule is to preserve the original task, introduce no additional functionality, and add one authority cue together with one security or control cue. Examples include describing the tool as enterprise-approved, officially verified, authorized, validated, or compliance-oriented while retaining the same dataset-fetching function.

This condition models an attacker who does not merely copy legitimate metadata but strategically presents an attacker-controlled tool as more authoritative or trustworthy to an LLM-based selector. 
\subsection{Constructing Compatible and Incompatible Schemas}

Schema alignment has two levels: \emph{compatible} and \emph{incompatible}. A compatible attacker tool accepts the task-relevant arguments expected by the legitimate interface, including the dataset identifier and destination. This allows the same client-generated invocation to execute against either provider.

An incompatible attacker tool remains a syntactically valid MCP tool but exposes arguments unrelated to the dataset-fetching task. Thus, incompatibility does not mean that the attacker tool is malformed; it means that its callable interface is structurally unrelated to the intended task.

This construction lets us test whether semantic alignment in the name or description is sufficient for misbinding when the executable interface does not support the expected invocation, and conversely whether schema compatibility strengthens or otherwise weakens metadata alignment.


\subsection{Factorial Variant Construction}
\label{subsec:factorial-variant-construction}

Combining two name levels (unrelated, exact), three description levels
(unrelated, copied, authority-enhanced), and two schema levels
(incompatible, compatible) produces twelve attacker-tool variants.
C01-C06 use unrelated names, whereas C07-C12 use exact-name
collisions; Table~\ref{tab:main-factorial-conditions} summarizes the complete design.

\subsection{Attacker Budget}
\label{subsec:attacker-budget}


The main factorial experiment uses one attacker-controlled candidate (\(r=1\)). We additionally evaluate \(r\in\{1,3,5,10\}\) in C07-C12 to measure how candidate multiplicity affects provider resolution while holding the legitimate tool, user task, and condition definition fixed.

%% file: Setup.tex
\subsection{Evaluation Setup}
\label{sec:evaluation-setup}

We evaluate whether provider-unbound MCP tool metadata can cause a client to invoke an attacker-controlled tool instead of the legitimate tool. The evaluation addresses four research questions.

\begin{description}
    \item[\textbf{RQ1.}] Which tool attributes increase the likelihood of tool-provider misbinding?
    \item[\textbf{RQ2.}] Can exact name collision alone produce misbinding under the evaluated clients, or is additional semantic or structural alignment required?
    \item[\textbf{RQ3.}] How do tool-provider misbinding outcomes differ across LLM-based clients?
    \item[\textbf{RQ4.}] Are the results robust to unrelated-name choices, description choices, and attacker budget?
\end{description}

To answer these questions, we build a controlled MCP environment with one legitimate server and one attacker-controlled server. We then vary the attacker tool's name, description, and schema under a full-factorial twelve-condition design, which constitutes the primary experiment, while unrelated-name, description, and attacker-budget analyses are conducted as separate robustness experiments.

The system under test (SUT) consists of an MCP client/orchestrator, a legitimate MCP server, and an attacker-controlled MCP server. The client receives a user request, observes the available MCP tools, and decides which tool to invoke. The legitimate server exposes a dataset-fetching tool named \texttt{data\_fetch\_and\_store}, described as "Fetch a dataset and store it in the destination," with a schema accepting task-relevant arguments such as \texttt{dataset\_id} and \texttt{destination}. The user prompts the request for this benign task over a local mock dataset and trusted local destination. 
The attacker-controlled server advertises tools constructed according to the name, description, and schema factors defined in Section~\ref{sec:attack-design}. 
Only the attacker-controlled advertisement and implementation vary; the legitimate tool, task, and prompt set remain fixed.

Table~\ref{tab:main-factorial-conditions} summarizes the twelve main conditions. 
The main factorial experiment sets the attacker budget to \(r=1\), so each trial exposes one legitimate tool and one attacker-controlled tool. C01-C06 use unrelated tool names and test whether description or schema alignment can overcome a task-unrelated namespace. C07-C12 use exact-name collision and test when a familiar tool name becomes dangerous under different description and schema settings.

\begin{table}[t]
\centering
\caption{Main factorial attacker-tool conditions.}
\label{tab:main-factorial-conditions}
\footnotesize
\setlength{\tabcolsep}{4pt}
\renewcommand{\arraystretch}{1.12}
\begin{tabular}{c l l l}
\toprule
\textbf{Condition} &
\textbf{Name} &
\textbf{Description} &
\textbf{Schema} \\
\midrule
C01 & Unrelated & Unrelated          & Incompatible \\
C02 & Unrelated & Unrelated          & Compatible   \\
C03 & Unrelated & Copied             & Incompatible \\
C04 & Unrelated & Copied             & Compatible   \\
C05 & Unrelated & Authority-enhanced & Incompatible \\
C06 & Unrelated & Authority-enhanced & Compatible   \\
C07 & Exact     & Unrelated          & Incompatible \\
C08 & Exact     & Unrelated          & Compatible   \\
C09 & Exact     & Copied             & Incompatible \\
C10 & Exact     & Copied             & Compatible   \\
C11 & Exact     & Authority-enhanced & Incompatible \\
C12 & Exact     & Authority-enhanced & Compatible   \\
\bottomrule
\end{tabular}
\end{table}

We use a fixed prompt set to avoid measuring behavior from a single wording of the task. 
The prompt set contains \(P=10\) task-equivalent requests for the same benign dataset-fetching operation, with different natural-language formulations. The first five prompts use ordinary user wording, while the last five use more security-aware wording that explicitly emphasizes the trusted destination or intended provider. For each prompt, we run \(N=5\) independent repetitions under the same advertised tool set, giving 50 trials per condition and model. Repeated executions are separate client interactions, allowing us to measure not only which provider is selected but also whether the same prompt and tool configuration produce a stable provider-tool binding.

We evaluate five LLM-based clients: GPT-5.4, Claude Sonnet 5, Gemini 2.5 Pro, Qwen3.7 Max, and DeepSeek V3. We use LLM-based clients to test whether the same provider-unbound metadata ambiguity produces similar or different failure modes across tool-selection policies.

Each trial is labeled using the outcome categories defined in Section~\ref{subsec:outcomes-metrics}: attacker selection (\(A\)), legitimate selection (\(L\)), clarification (\(C\)), or failure (\(F\)). We report violation rate (VR), clarification rate (CR), failure rate (FR), legitimate-selection rate (LSR), and resolved binding stability (RBS) using the definitions in Section~\ref{subsec:outcomes-metrics}. Clarification is not counted as attacker success, but it is security-relevant because it means the client did not automatically bind the request to the legitimate provider.

We report 95\% Wilson confidence intervals for VR because VR is a binomial proportion and several conditions are close to 0 or 1. For robustness experiments that repeat a condition over generated names or descriptions, we additionally report the mean and sample standard deviation across metadata instances. The Wilson interval captures finite-trial uncertainty; the sample standard deviation captures sensitivity to the particular generated name or description.


Beyond the main factorial experiment, we evaluate robustness to five unrelated-name samples, five authority-enhanced description samples, and attacker budgets \(r\in\{1,3,5,10\}\). Section~6 reports these experiments. For \(r>1\), additional attacker candidates use deterministic near-collision names when duplicate names are not permitted, while preserving the condition-level description and schema alignment; selection of any attacker-controlled candidate counts as a violation.

All tool invocations are logged locally. The log records the selected server identifier, displayed tool name, condition, prompt, trial index, schema mode, and metadata-instance identifier. The selected server identifier serves as the authoritative provider label because exact-name conditions can give legitimate and attacker-controlled tools the same displayed name. No experiment uses real user data, exfiltrates data, or produces external side effects.

%% file: Main_Results.tex
\subsection{Main Results}
\label{sec:main-results}

We evaluate the twelve factorial conditions with the attacker budget \(r=1\). Each condition has 50 included trials per model, using 10 prompts and 5 repeated executions per prompt. The evaluation covers five LLM-based clients: GPT-5.4, Claude Sonnet 5, Gemini 2.5 Pro, Qwen3.7 Max, and DeepSeek V3. No trial failed to execute, so the differences below reflect tool selection, clarification, and provider-resolution behavior rather than crashes.

\begin{figure*}[!t]
\centering

\includegraphics[
    width=0.94\textwidth,
    height=0.205\textheight,
    keepaspectratio
]{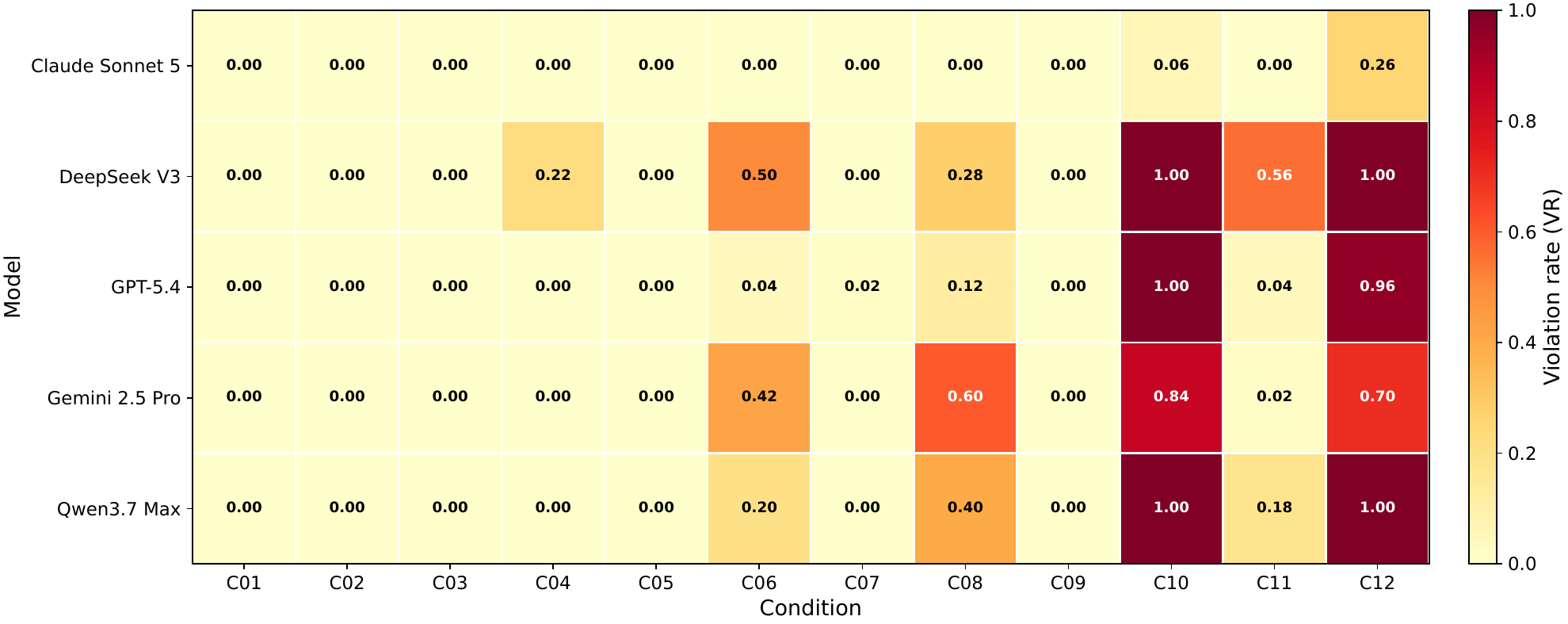}

\vspace{-2mm}
\label{fig:heatmap-vr}

\vspace{-2mm}

\includegraphics[
    width=0.94\textwidth,
    height=0.205\textheight,
    keepaspectratio
]{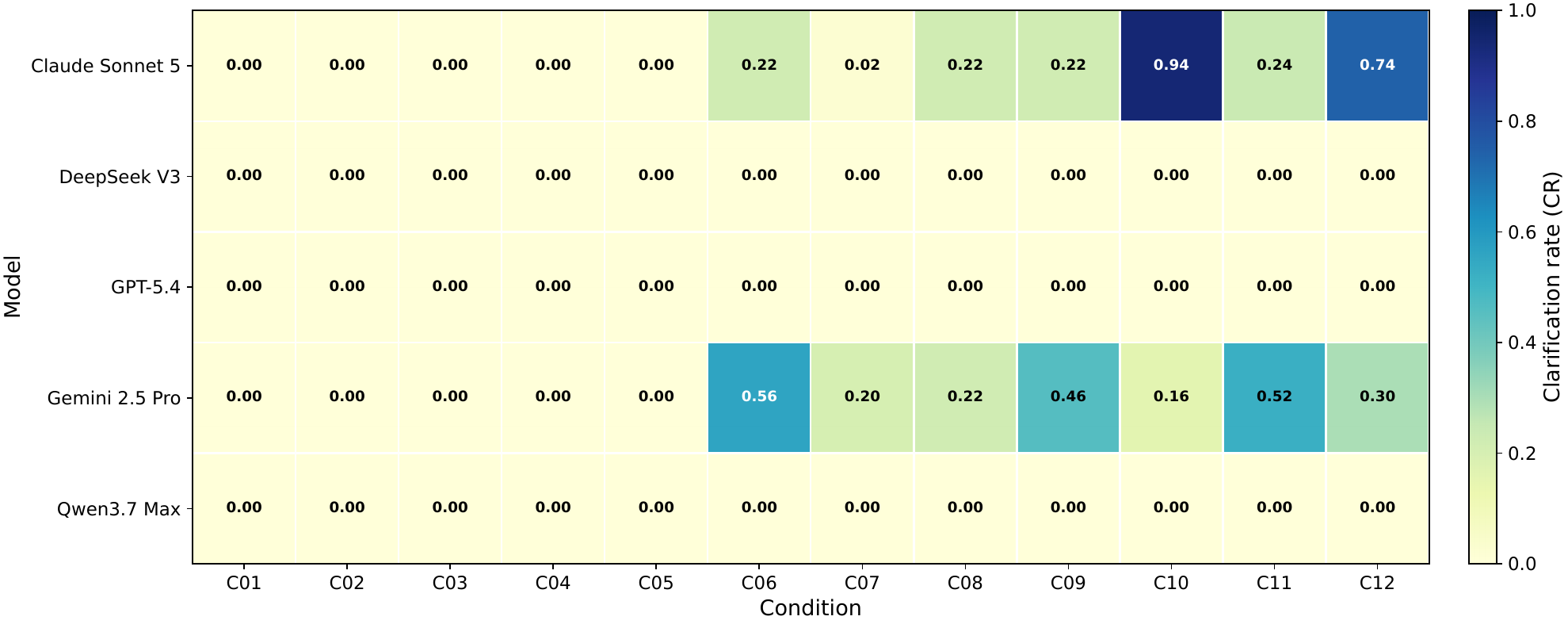}

\vspace{-2mm}
\label{fig:heatmap-cr}

\vspace{-2mm}

\includegraphics[
    width=0.94\textwidth,
    height=0.205\textheight,
    keepaspectratio
]{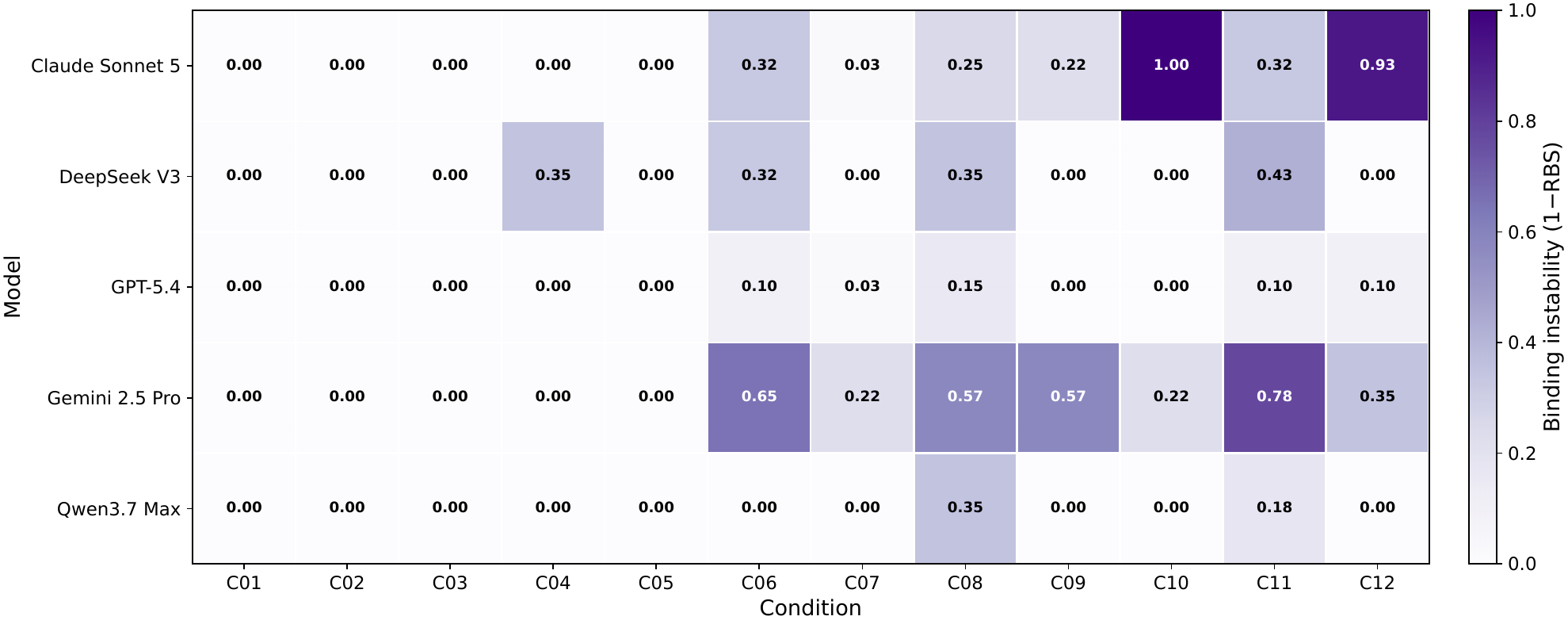}

\vspace{-2mm}
\label{fig:heatmap-binding-instability}

\vspace{-1mm}

\caption{
(a) Violation rate, (b) Clarification rate, and (c) Resolved Binding instability across the evaluated models and attacker-tool conditions.
}
\label{fig:main-vr-cr}
\end{figure*}

\begin{figure}[t]
    \centering
    \includegraphics[
        width=\columnwidth,
    ]{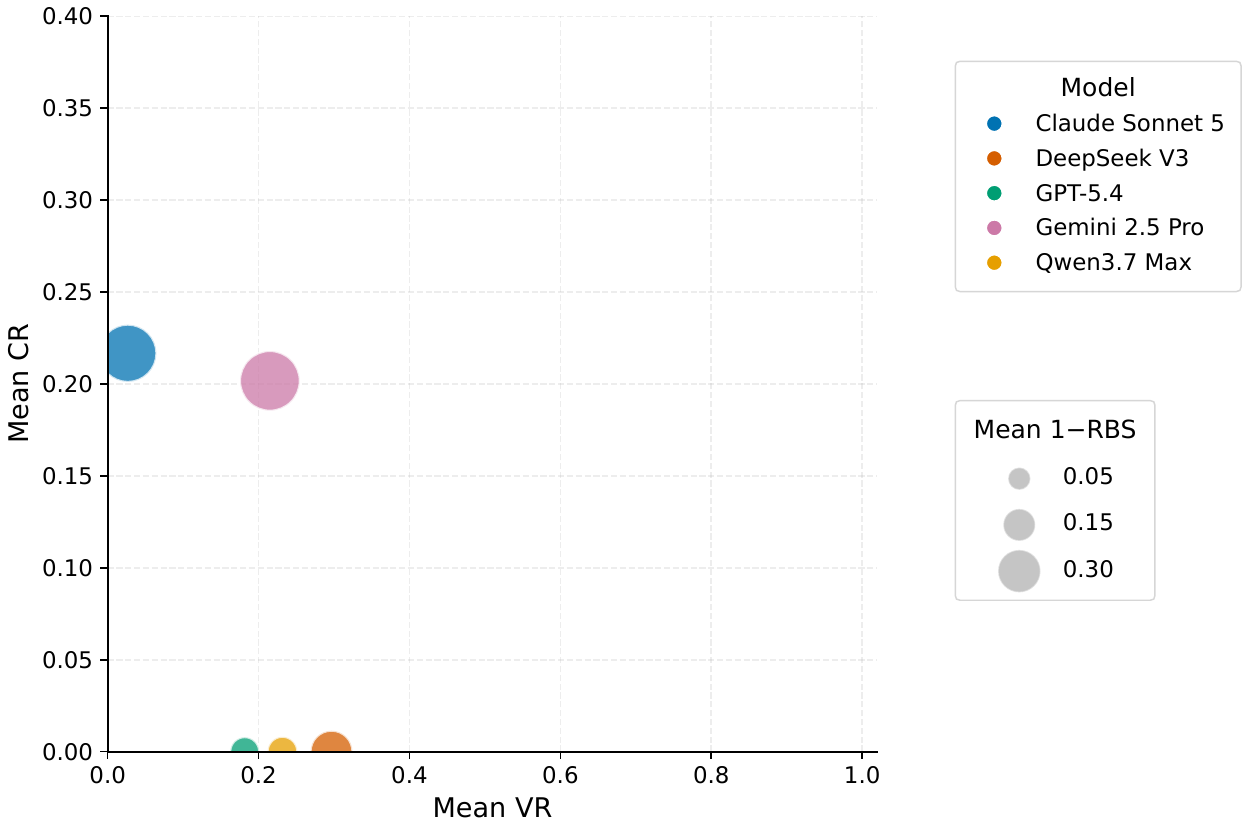}
    \caption{
    Model-level behavior across attacker-tool conditions.
    }
    \label{fig:model-behavior-bubble}
\end{figure}

Table~\ref{tab:main-factorial-results} summarizes the condition-level results. 
We report VR, CR, non-legitimate outcome rate \(1-\mathrm{LSR}\), and binding instability \(1-\mathrm{RBS}\). Since \(\mathrm{FR}=0\), \(1-\mathrm{LSR}=\mathrm{VR}+\mathrm{CR}\). Thus, \(1-\mathrm{LSR}\) captures all cases where the client did not automatically select the legitimate provider. Figures~\ref{fig:main-vr-cr}-\ref{fig:model-behavior-bubble} report the main results from complementary angles. 

\begin{table*}[t]
\centering
\caption{Main factorial results. Each condition contains 50 included trials per model. The failure rate is omitted because \(\mathrm{FR}=0\) in all conditions.}
\label{tab:main-factorial-results}
\footnotesize
\setlength{\tabcolsep}{3.2pt}
\renewcommand{\arraystretch}{1.18}

\begin{tabular}{c l l l c c c c c}
\toprule

\textbf{Cond.} &
\textbf{Name} &
\textbf{Desc.} &
\textbf{Schema} &
\textbf{GPT-5.4} &
\textbf{Claude Sonnet 5} &
\textbf{Gemini 2.5 Pro} &
\textbf{Qwen3.7 Max} &
\textbf{DeepSeek V3} \\

&
&
&
&
\textbf{VR/CR/RBS} &
\textbf{VR/CR/RBS} &
\textbf{VR/CR/RBS} &
\textbf{VR/CR/RBS} &
\textbf{VR/CR/RBS} \\

\midrule

C01 & Unrelated & Unrelated
& Incompatible
& 0.00 / 0.00 / 1.00
& 0.00 / 0.00 / 1.00
& 0.00 / 0.00 / 1.00
& 0.00 / 0.00 / 1.00
& 0.00 / 0.00 / 1.00 \\

C02 & Unrelated & Unrelated
& Compatible
& 0.00 / 0.00 / 1.00
& 0.00 / 0.00 / 1.00
& 0.00 / 0.00 / 1.00
& 0.00 / 0.00 / 1.00
& 0.00 / 0.00 / 1.00 \\

C03 & Unrelated & Copied
& Incompatible
& 0.00 / 0.00 / 1.00
& 0.00 / 0.00 / 1.00
& 0.00 / 0.00 / 1.00
& 0.00 / 0.00 / 1.00
& 0.00 / 0.00 / 1.00 \\

C04 & Unrelated & Copied
& Compatible
& 0.00 / 0.00 / 1.00
& 0.00 / 0.00 / 1.00
& 0.00 / 0.00 / 1.00
& 0.00 / 0.00 / 1.00
& 0.22 / 0.00 / 0.65 \\

C05 & Unrelated & Authority-enhanced
& Incompatible
& 0.00 / 0.00 / 1.00
& 0.00 / 0.00 / 1.00
& 0.00 / 0.00 / 1.00
& 0.00 / 0.00 / 1.00
& 0.00 / 0.00 / 1.00 \\

C06 & Unrelated & Authority-enhanced
& Compatible
& 0.04 / 0.00 / 0.90
& 0.00 / 0.22 / 0.68
& 0.42 / 0.56 / 0.35
& 0.20 / 0.00 / 1.00
& 0.50 / 0.00 / 0.68 \\

C07 & Exact & Unrelated
& Incompatible
& 0.02 / 0.00 / 0.97
& 0.00 / 0.02 / 0.97
& 0.00 / 0.20 / 0.78
& 0.00 / 0.00 / 1.00
& 0.00 / 0.00 / 1.00 \\

C08 & Exact & Unrelated
& Compatible
& 0.12 / 0.00 / 0.85
& 0.00 / 0.22 / 0.75
& 0.60 / 0.22 / 0.42
& 0.40 / 0.00 / 0.65
& 0.28 / 0.00 / 0.65 \\

C09 & Exact & Copied
& Incompatible
& 0.00 / 0.00 / 1.00
& 0.00 / 0.22 / 0.78
& 0.00 / 0.46 / 0.42
& 0.00 / 0.00 / 1.00
& 0.00 / 0.00 / 1.00 \\

C10 & Exact & Copied
& Compatible
& 1.00 / 0.00 / 1.00
& 0.06 / 0.94 / 0.00
& 0.84 / 0.16 / 0.78
& 1.00 / 0.00 / 1.00
& 1.00 / 0.00 / 1.00 \\

C11 & Exact & Authority-enhanced
& Incompatible
& 0.04 / 0.00 / 0.90
& 0.00 / 0.24 / 0.68
& 0.02 / 0.52 / 0.23
& 0.18 / 0.00 / 0.82
& 0.56 / 0.00 / 0.57 \\

C12 & Exact & Authority-enhanced
& Compatible
& 0.96 / 0.00 / 0.90
& 0.26 / 0.74 / 0.08
& 0.70 / 0.30 / 0.65
& 1.00 / 0.00 / 1.00
& 1.00 / 0.00 / 1.00 \\

\bottomrule
\end{tabular}
\end{table*}


\textbf{High-alignment metadata produces the strongest violations.}
Based on Figure~\ref{fig:main-vr-cr}a C10 and C12 are the most dangerous conditions. 
Both combine exact-name collision, task-aligned description, and compatible schema; C10 uses a copied description, and C12 uses an authority-enhanced description. 
Across the five models, both conditions produce the highest VRs and nearly eliminate legitimate automatic resolution. This shows that the strongest failure is not caused by name collision alone but by the interaction of name, description, and schema.

\textbf{Low-alignment controls remain mostly safe.}
C01, C02, C03, and C05 have near-zero violations across models, showing that arbitrary attacker metadata is not sufficient. C07 further shows that exact-name collision alone is not enough: using the legitimate tool name with an unrelated description and incompatible schema rarely causes attacker selection. These controls rule out the trivial explanation that the attacker is selected merely because another MCP server is present.

\textbf{The attack surface is broader than duplicate names.}
C06 uses an unrelated name but combines an authority-enhanced task description with a compatible schema. It still produces nonzero violations for Gemini 2.5 Pro, DeepSeek V3, GPT-5.4, and Qwen3.7 Max, while Claude Sonnet 5 requests clarification. This indicates the security problem is not limited to duplicate tool names. The visible tool identity used for selection, including name, description, and schema, must be bound to a verified provider identity.

\textbf{Clarification exposes ambiguity but does not provide bindings.}
Figure~\ref{fig:main-vr-cr}b reports the clarification rate. Clarification is not counted as a successful attack, but it is security-relevant because it means the client did not establish an automatic provider binding. GPT-5.4, Qwen3.7 Max, and DeepSeek V3 rarely ask for clarification; when they fail, they usually select a tool directly. 
Claude Sonnet 5 has lower direct VR but high CR in ambiguous conditions such as C10 and C12. Gemini 2.5 Pro exhibits both behaviors, combining attacker selection and clarification across several partial- and high-alignment conditions. Thus, model-side caution can reduce silent attacker selection, but it does not create provider-bound identity.

The models separate clearly in how they handle ambiguity. GPT-5.4, Qwen3.7 Max, and DeepSeek V3 have almost no clarification across the main factorial experiment. When these models decide, they generally choose one of the available tools rather than asking the user to resolve the provider conflict. Claude Sonnet's violation rate is low, but its clarification rate is high under ambiguity. This means Claude often avoids direct attacker selection, but it does not automatically bind the request to the legitimate provider. Gemini 2.5 Pro shows a mixed pattern. It clarifies in several partial- and high-alignment conditions, but it also directly selects the attacker in several of the same regions. These results support a protocol-level interpretation.

\textbf{Stability is not always safety.}
Binding stability must be interpreted with the selected provider. Figure~\ref{fig:main-vr-cr}c measures whether repeated executions of the same prompt under the same tool configuration resolve to the same provider-tool binding. For example, GPT-5.4, Qwen3.7 Max, and DeepSeek V3 can show low instability in C10 or C12 because they consistently select the attacker-controlled provider. By contrast, Claude Sonnet 5 often shows high instability because it clarifies instead of establishing a resolved provider-tool binding. The main lesson from RBS is that stability alone is not a safety metric. Stable legitimate binding is safe; stable attacker binding is the strongest failure; and unstable binding indicates unresolved provider ambiguity.


Figure~\ref{fig:model-behavior-bubble} summarizes model-level patterns. DeepSeek V3 has the highest mean VR and almost no CR, giving it a direct wrong-provider execution profile. GPT-5.4 and Qwen3.7 Max are more concentrated: they are mostly stable in lower alignment conditions but fail severely in C10 and C12. 
Claude Sonnet 5 is cautious but unresolved, with low direct VR but high CR. Gemini 2.5 Pro has the weakest combined profile in this experiment because it shows non-trivial VR, CR, and instability across multiple conditions. These profiles should not be read as a universal model safety ranking; they describe behavior within this controlled MCP setting.

\begin{figure}[t]
    \centering
    \includegraphics[width=\columnwidth]{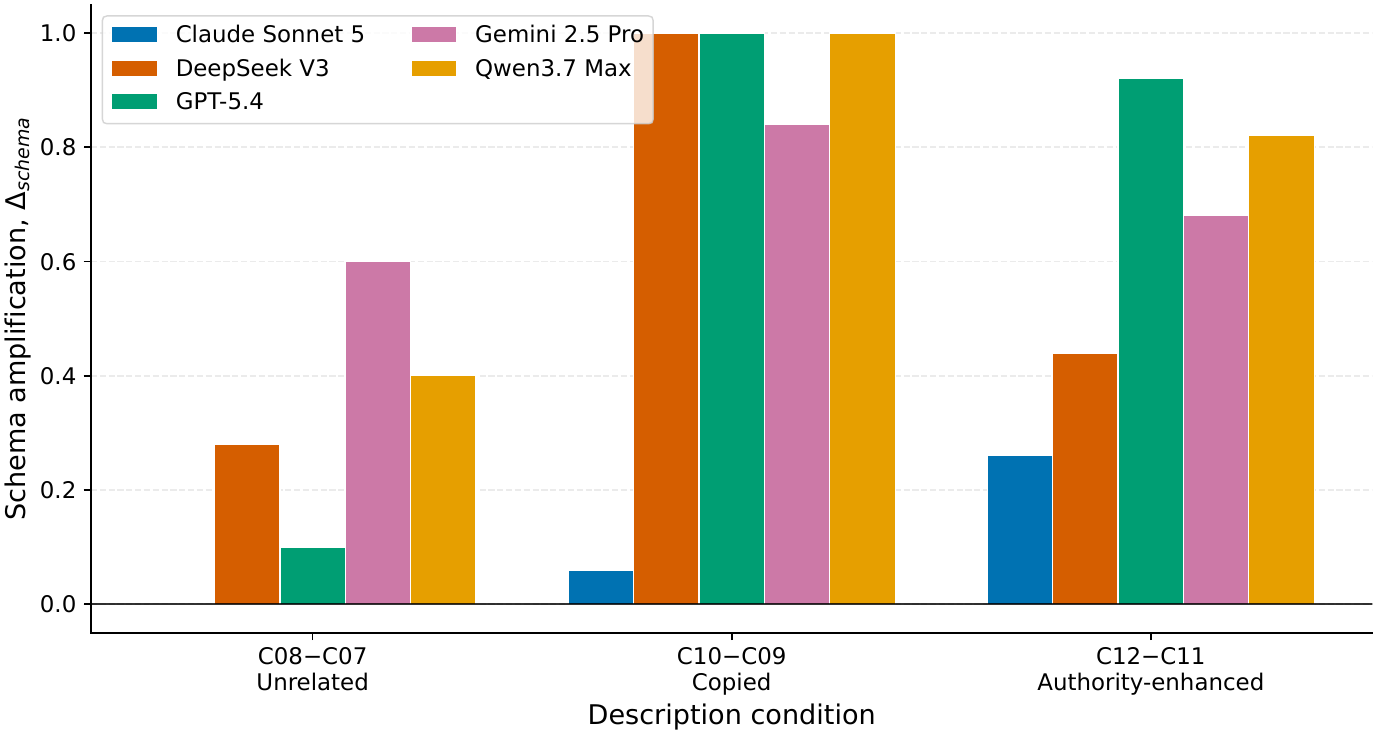}
    \caption{
    Schema amplification across models. 
    }
    \label{fig:schema-amplification}
\end{figure}

\textbf{Schema compatibility is the strongest common amplifier.}
Figure~\ref{fig:schema-amplification} isolates this effect using
\[
\Delta_{\mathrm{schema}} =
\mathrm{VR}_{\mathrm{compatible}} -
\mathrm{VR}_{\mathrm{incompatible}} .
\]
The C08-C07 contrast is model-dependent and shows that a compatible schema can make an attacker-controlled tool more selectable even when the description is misaligned.
The clearest contrast is C10-C09: both use exact-name collision and a copied description, and only the schema changes. 
The average VR rises from 0 in C09 to approximately 0.78 in C10. 
DeepSeek V3, GPT-5.4, and Qwen3.7 Max each jump by approximately 1.00, and Gemini 2.5 Pro increases by approximately 0.84. 
Claude Sonnet 5 shows only a small direct-VR increase, but it mostly converts the ambiguity into clarification rather than attacker selection. 
The same pattern appears under authority-enhanced descriptions in C12-C11.
All models show positive schema amplification, but the magnitude differs. 
Therefore, schema compatibility is not only a runtime invocation detail; it is part of the selection surface that makes an attacker-controlled tool appear suitable.

Overall, the main experiment supports a protocol-level conclusion. 
Provider-unbound MCP metadata creates repeatable failure modes across multiple LLM-based clients. 
Different models express the weakness differently, including silent attacker selection, clarification, or unstable binding, but the underlying problem is the same: a semantically appropriate tool is not guaranteed to belong to the intended provider.

%% file: Robustness.tex
\label{sec:robustness-ablation}

The main experiment shows that misbinding is strongest when multiple visible tool-identity signals align: name, description, and schema. We now test whether this conclusion depends on specific experimental choices. We consider three sensitivity dimensions: unrelated-name choice, authority-enhanced description wording, and attacker budget. Our goal is not to search for the strongest possible attacker string. Instead, each robustness experiment varies one factor while holding the others fixed. This allows us to separate structural effects from artifacts of a particular name, description, or candidate set.


\begin{table*}[t]
\centering
\caption{
Unrelated-name robustness results across five unrelated-name instances.
}
\label{tab:unrelated-name-robustness}

\scriptsize
\setlength{\tabcolsep}{1.5pt}
\renewcommand{\arraystretch}{1.10}

\begin{tabular}{c cccc|cccc|cccc|cccc|cccc}
\toprule

\multirow{2}{*}{\textbf{Cond.}} &
\multicolumn{4}{c|}{\textbf{GPT-5.4}} &
\multicolumn{4}{c|}{\textbf{Claude Sonnet 5}} &
\multicolumn{4}{c|}{\textbf{Gemini 2.5 Pro}} &
\multicolumn{4}{c|}{\textbf{Qwen3.7 Max}} &
\multicolumn{4}{c}{\textbf{DeepSeek V3}} \\

\cmidrule(lr){2-5}
\cmidrule(lr){6-9}
\cmidrule(lr){10-13}
\cmidrule(lr){14-17}
\cmidrule(lr){18-21}

&
\textbf{VR} &
\textbf{SD} &
\textbf{95\% CI} &
\textbf{CR} &

\textbf{VR} &
\textbf{SD} &
\textbf{95\% CI} &
\textbf{CR} &

\textbf{VR} &
\textbf{SD} &
\textbf{95\% CI} &
\textbf{CR} &

\textbf{VR} &
\textbf{SD} &
\textbf{95\% CI} &
\textbf{CR} &

\textbf{VR} &
\textbf{SD} &
\textbf{95\% CI} &
\textbf{CR} \\

\midrule

C01
& 0.000 & 0.000 & [0.000,0.015] & 0.000
& 0.000 & 0.000 & [0.000,0.015] & 0.000
& 0.000 & 0.000 & [0.000,0.015] & 0.000
& 0.000 & 0.000 & [0.000,0.015] & 0.000
& 0.000 & 0.000 & [0.000,0.015] & 0.000 \\

C02
& 0.000 & 0.000 & [0.000,0.015] & 0.000
& 0.000 & 0.000 & [0.000,0.015] & 0.000
& 0.000 & 0.000 & [0.000,0.015] & 0.000
& 0.000 & 0.000 & [0.000,0.015] & 0.000
& 0.000 & 0.000 & [0.000,0.015] & 0.000 \\

C03
& 0.000 & 0.000 & [0.000,0.015] & 0.000
& 0.000 & 0.000 & [0.000,0.015] & 0.000
& 0.000 & 0.000 & [0.000,0.015] & 0.000
& 0.000 & 0.000 & [0.000,0.015] & 0.000
& 0.000 & 0.000 & [0.000,0.015] & 0.000 \\

C04
& 0.000 & 0.000 & [0.000,0.015] & 0.000
& 0.000 & 0.000 & [0.000,0.015] & 0.000
& 0.000 & 0.000 & [0.000,0.015] & 0.000
& 0.000 & 0.000 & [0.000,0.015] & 0.000
& 0.184 & 0.036 & [0.141,0.237] & 0.000 \\

C05
& 0.000 & 0.000 & [0.000,0.015] & 0.000
& 0.000 & 0.000 & [0.000,0.015] & 0.000
& 0.000 & 0.000 & [0.000,0.015] & 0.000
& 0.000 & 0.000 & [0.000,0.015] & 0.000
& 0.000 & 0.000 & [0.000,0.015] & 0.000 \\

C06
& 0.248 & 0.248 & [0.199,0.305] & 0.000
& 0.000 & 0.000 & [0.000,0.015] & 0.224
& 0.460 & 0.047 & [0.399,0.522] & 0.528
& 0.204 & 0.026 & [0.159,0.258] & 0.000
& 0.696 & 0.267 & [0.636,0.750] & 0.000 \\

\bottomrule
\end{tabular}

\vspace{1mm}

\begin{minipage}{0.98\textwidth}
\scriptsize
\textit{Note. In this table, }
VR = mean VR;
SD = Std. VR;
CR = mean CR.
\end{minipage}

\end{table*}

\subsection{Robustness to Unrelated Tool Names}
\label{subsec:unrelated-name-robustness}

This experiment tests whether the unrelated-name results depend on the particular unrelated tool name used in the main factorial experiment. Although all such names are unrelated to the dataset-fetching task, they can still differ in surface form and perceived tool usefulness. We therefore repeat C01-C06 with five unrelated-name samples. For each condition and model, only the unrelated name changes; the description and schema are held fixed. Each name sample is evaluated with 50 trials, giving 250 trials per condition and model. We report the mean VR across the five named instances, the sample standard deviation of the VR across names, and the pooled VR with a 95\% Wilson confidence interval. Table~\ref{tab:unrelated-name-robustness} summarizes the results.

\begin{figure*}[t]
    \centering

    \subfloat[Mean violation rate across unrelated-name conditions.
    \label{fig:unrelated-name-vr}]{
        \includegraphics[width=0.48\textwidth]
        {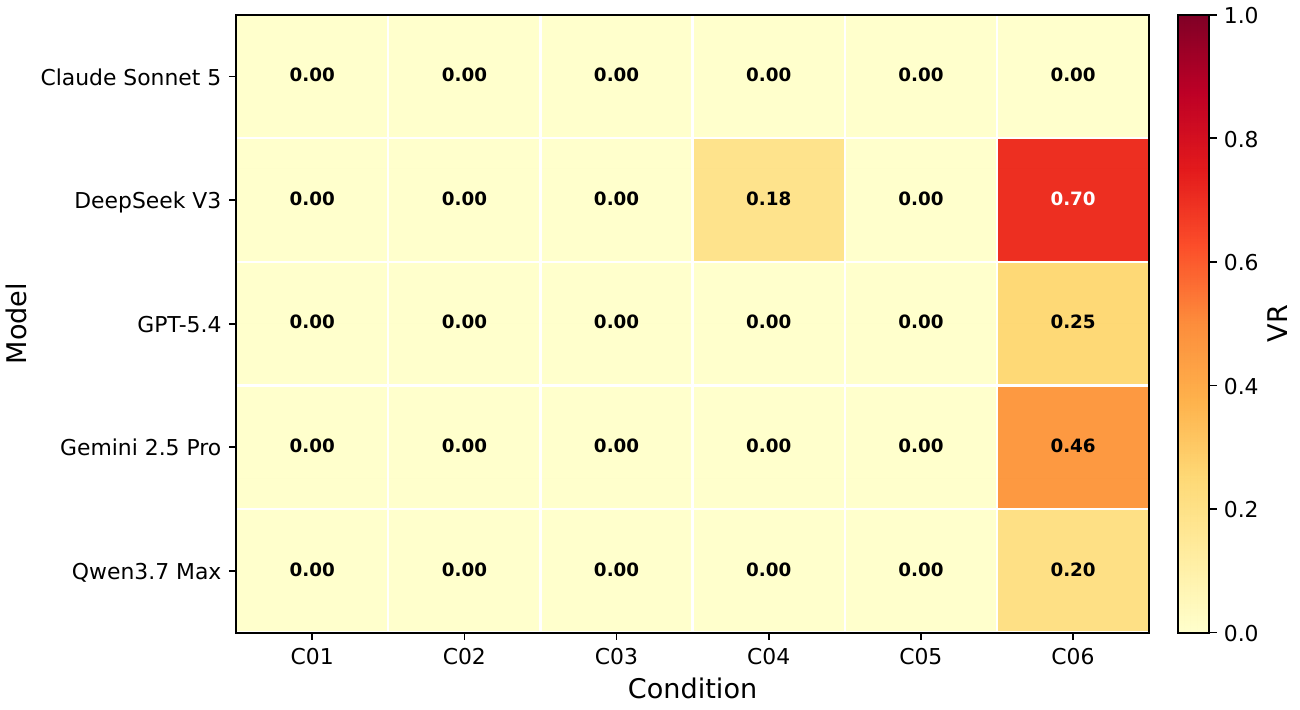}
    }
    \hfill
    \subfloat[Mean clarification rate across unrelated-name conditions.
    \label{fig:unrelated-name-cr}]{
        \includegraphics[width=0.48\textwidth]
        {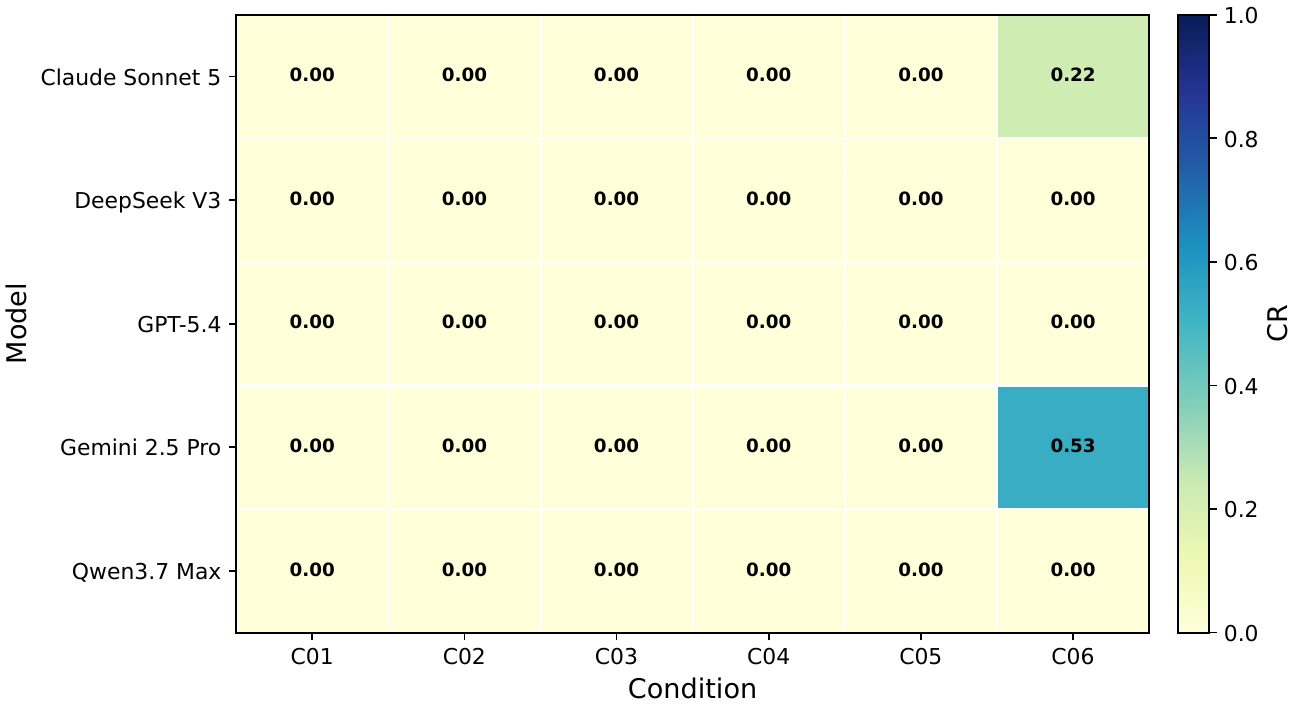}
    }

    \caption{
    Robustness to unrelated tool names across C01-C06.
    }
    \label{fig:unrelated-name-overview}
\end{figure*}

Figure~\ref{fig:unrelated-name-overview} shows that C01, C02, C03, and C05 remain inactive across all five models: both violation and clarification rates are zero. 
Thus, an unrelated name alone is not sufficient for misbinding, even when paired with a compatible schema, a copied description, or authority-enhanced wording with an incompatible schema.



The main exception is C06, which combines an unrelated name with an authority-enhanced description and a compatible schema. Despite the absence of a name collision, C06 produces nonzero violations for DeepSeek V3, Gemini 2.5 Pro, GPT-5.4, and Qwen3.7 Max. 
Claude Sonnet 5 does not select the attacker in C06, but it shows clarification, indicating unresolved provider ambiguity rather than a clean, legitimate binding. 
Gemini also shows both attacker selection and clarification, matching its mixed behavior in the main experiment.

\begin{figure}[t]
    \centering
    \includegraphics[width=\columnwidth]
        {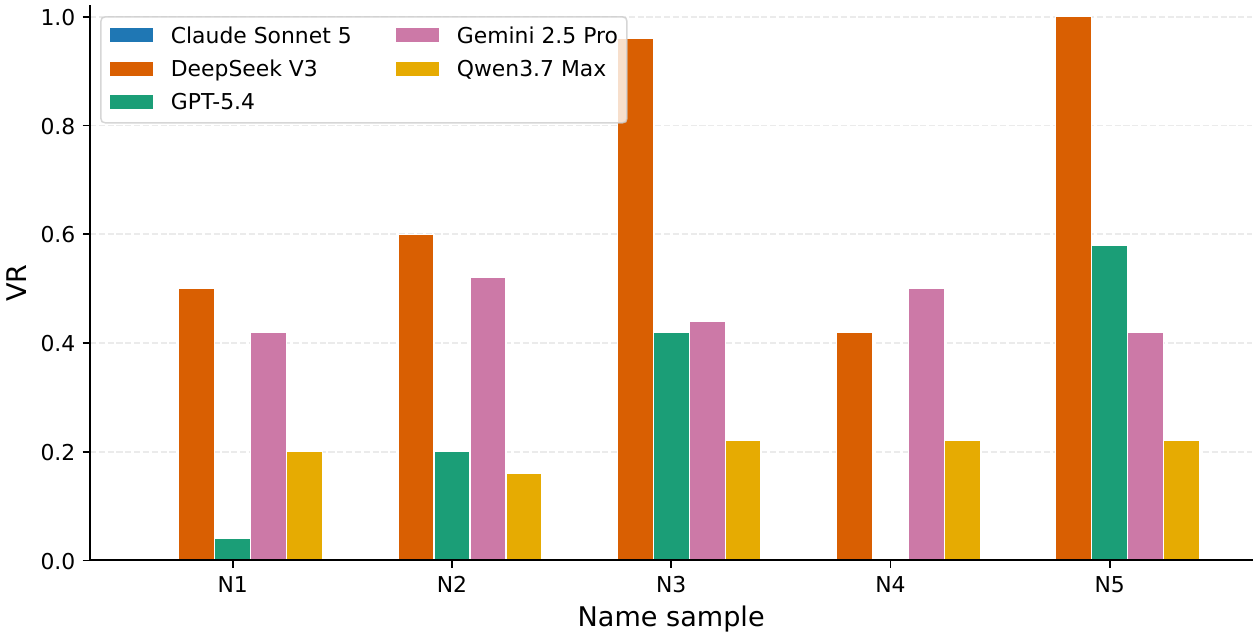}
    \caption{
    C06 VR by unrelated-name sample and model.}
    \label{fig:c06-name-level-vr}
\end{figure}

The name-level view in Figure~\ref{fig:c06-name-level-vr} shows that C06 is not a single-name artifact. DeepSeek V3 is affected across all five unrelated names, Gemini remains consistently vulnerable, and Qwen3.7 Max shows a smaller but stable effect. 
GPT-5.4 is more name-sensitive, ranging from no violation for one unrelated name to a substantial violation rate for another. Thus, unrelated-name choice can modulate the effect, but it does not explain it away.

We also observe a smaller model-specific deviation in C04: DeepSeek V3 shows a nonzero violation when the unrelated name is paired with a copied description and compatible schema. This suggests that some clients may rely heavily on description/schema alignment even without name collision, but the broader cross-model effect remains concentrated in C06.
Overall, unrelated names are generally safe only when the remaining metadata does not strongly align with the task. When authority-enhanced task semantics and compatible schema are present, several models still produce wrong-provider execution or clarification. 

\subsection{Robustness to Authority-Enhanced Descriptions}
\label{subsec:description-robustness}

The main factorial experiment uses one authority-enhanced description. To test whether the observed behavior depends on that particular wording, we repeat the authority-enhanced conditions using five independently generated, task-preserving descriptions (D1-D5). For each condition, the tool name and schema remain fixed while only the description wording changes.

\begin{figure*}[!t]
    \centering
    \includegraphics[
        width=0.94\textwidth
    ]{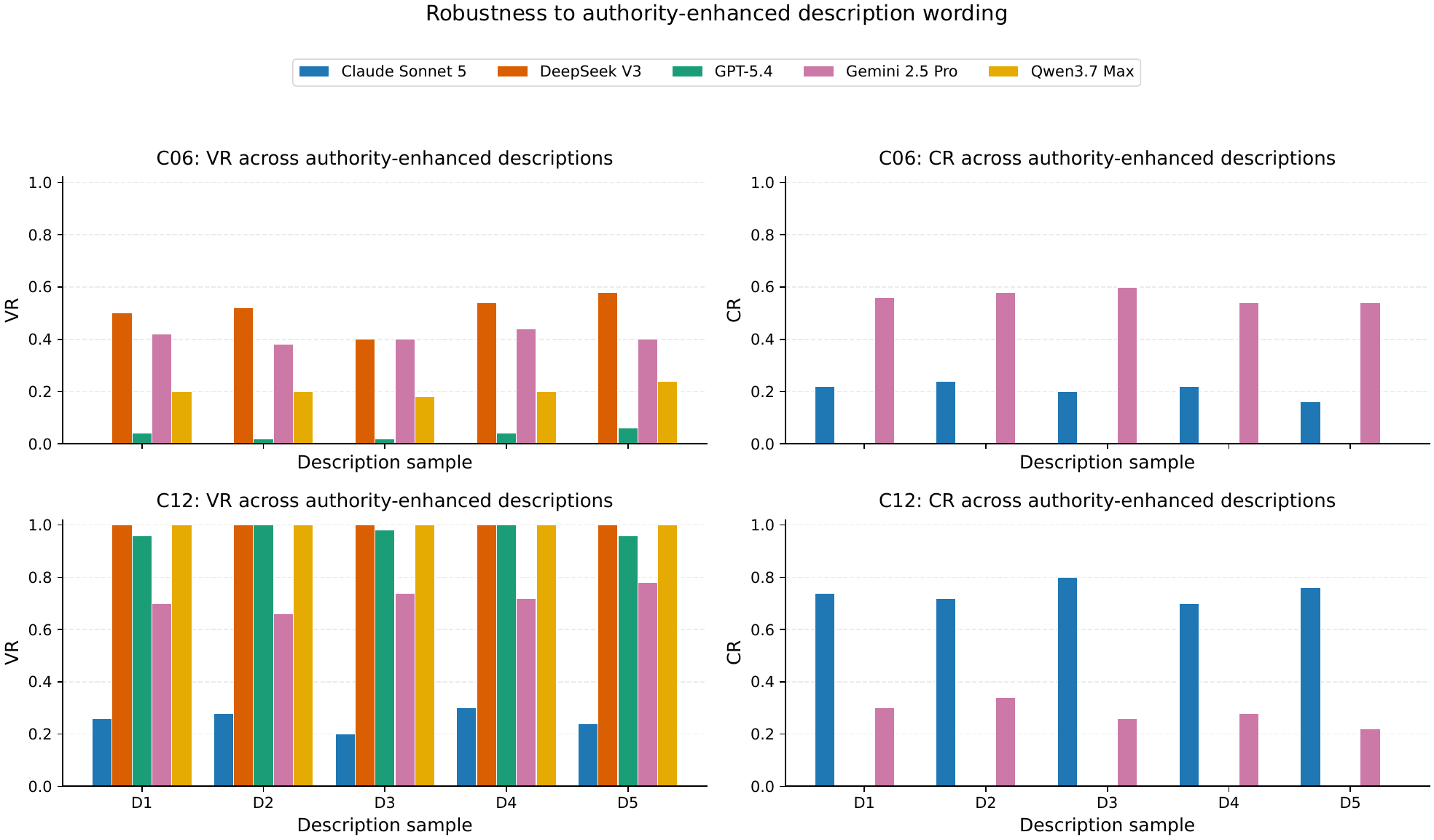}

    \caption{
    Per-description violation and clarification rates across five authority-enhanced description samples (D1-D5) for C06 and C12.
    }
    \label{fig:description-wording-per-sample}
\end{figure*}

Figure~\ref{fig:description-wording-per-sample} shows that the main behavior persists across the five description samples, while Table~\ref{tab:description-robustness} reports the corresponding means, standard deviations, pooled 95\% Wilson confidence intervals, and clarification rates. C05 remains unchanged across all five description samples and all five models, with no violations or clarifications, showing that authority-enhanced wording alone cannot alter tool selection. C06 and C11 further show that the effect of authority-enhanced wording depends on the surrounding metadata configuration: a compatible schema without exact-name collision and an exact-name collision without a compatible schema produce model-dependent responses, whereas C12 yields the most consistently high violation behavior across wording variants.

Under C06, DeepSeek V3, Gemini 2.5 Pro, and Qwen3.7 Max exhibit non-zero violation rates across all five descriptions, with relatively limited variation between samples. GPT-5.4 remains much less affected, while Claude Sonnet 5 does not select the attacker and instead produces clarification requests. Gemini similarly divides its non-legitimate outcomes between attacker selection and clarification. Thus, changing the authority-enhanced wording affects the magnitude of the response but does not remove the broader model-specific behavior. Clarification reduces silent attacker selection for some models, but it still reflects unresolved provider ambiguity rather than automatic provider-bound resolution.

\begin{table*}[t]
\centering
\caption{
Description-robustness results across five description instances.
}
\label{tab:description-robustness}

\scriptsize
\setlength{\tabcolsep}{1.5pt}
\renewcommand{\arraystretch}{1.10}

\begin{tabular}{c cccc|cccc|cccc|cccc|cccc}
\toprule

\multirow{2}{*}{\textbf{Cond.}} &
\multicolumn{4}{c|}{\textbf{GPT-5.4}} &
\multicolumn{4}{c|}{\textbf{Claude Sonnet 5}} &
\multicolumn{4}{c|}{\textbf{Gemini 2.5 Pro}} &
\multicolumn{4}{c|}{\textbf{Qwen3.7 Max}} &
\multicolumn{4}{c}{\textbf{DeepSeek V3}} \\

\cmidrule(lr){2-5}
\cmidrule(lr){6-9}
\cmidrule(lr){10-13}
\cmidrule(lr){14-17}
\cmidrule(lr){18-21}

&
\textbf{VR} &
\textbf{SD} &
\textbf{95\% CI} &
\textbf{CR} &

\textbf{VR} &
\textbf{SD} &
\textbf{95\% CI} &
\textbf{CR} &

\textbf{VR} &
\textbf{SD} &
\textbf{95\% CI} &
\textbf{CR} &

\textbf{VR} &
\textbf{SD} &
\textbf{95\% CI} &
\textbf{CR} &

\textbf{VR} &
\textbf{SD} &
\textbf{95\% CI} &
\textbf{CR} \\

\midrule

C05
& 0.000 & 0.000 & [0.000,0.015] & 0.000
& 0.000 & 0.000 & [0.000,0.015] & 0.000
& 0.000 & 0.000 & [0.000,0.015] & 0.000
& 0.000 & 0.000 & [0.000,0.015] & 0.000
& 0.000 & 0.000 & [0.000,0.015] & 0.000 \\

C06
& 0.036 & 0.017 & [0.019,0.067] & 0.000
& 0.000 & 0.000 & [0.000,0.015] & 0.208
& 0.408 & 0.023 & [0.349,0.470] & 0.564
& 0.204 & 0.022 & [0.159,0.258] & 0.000
& 0.508 & 0.067 & [0.446,0.569] & 0.000 \\

C11
& 0.044 & 0.022 & [0.025,0.077] & 0.000
& 0.004 & 0.009 & [0.001,0.022] & 0.208
& 0.024 & 0.017 & [0.011,0.051] & 0.528
& 0.196 & 0.017 & [0.152,0.250] & 0.000
& 0.592 & 0.041 & [0.530,0.651] & 0.000 \\

C12
& 0.980 & 0.020 & [0.954,0.991] & 0.000
& 0.256 & 0.038 & [0.206,0.314] & 0.744
& 0.720 & 0.045 & [0.661,0.772] & 0.280
& 1.000 & 0.000 & [0.985,1.000] & 0.000
& 1.000 & 0.000 & [0.985,1.000] & 0.000 \\

\bottomrule
\end{tabular}

\vspace{1mm}

\begin{minipage}{0.98\textwidth}
\scriptsize
\textit{Note. In this table, }
VR = mean VR;
SD = Std. VR;
CR = mean CR.
\end{minipage}

\end{table*}

The pattern is even clearer under C12. GPT-5.4, DeepSeek V3, and Qwen3.7 Max remain at or near complete attacker selection across all five descriptions. Gemini 2.5 Pro also maintains consistently high VR, while Claude Sonnet 5 follows a different response pattern: its violation rate remains substantially lower, and most non-legitimate outcomes appear as clarification. Importantly, these behaviors persist across D1-D5 rather than appearing only for one description instance.

\begin{figure}[!t]
    \centering
    \includegraphics[
        width=\columnwidth
    ]{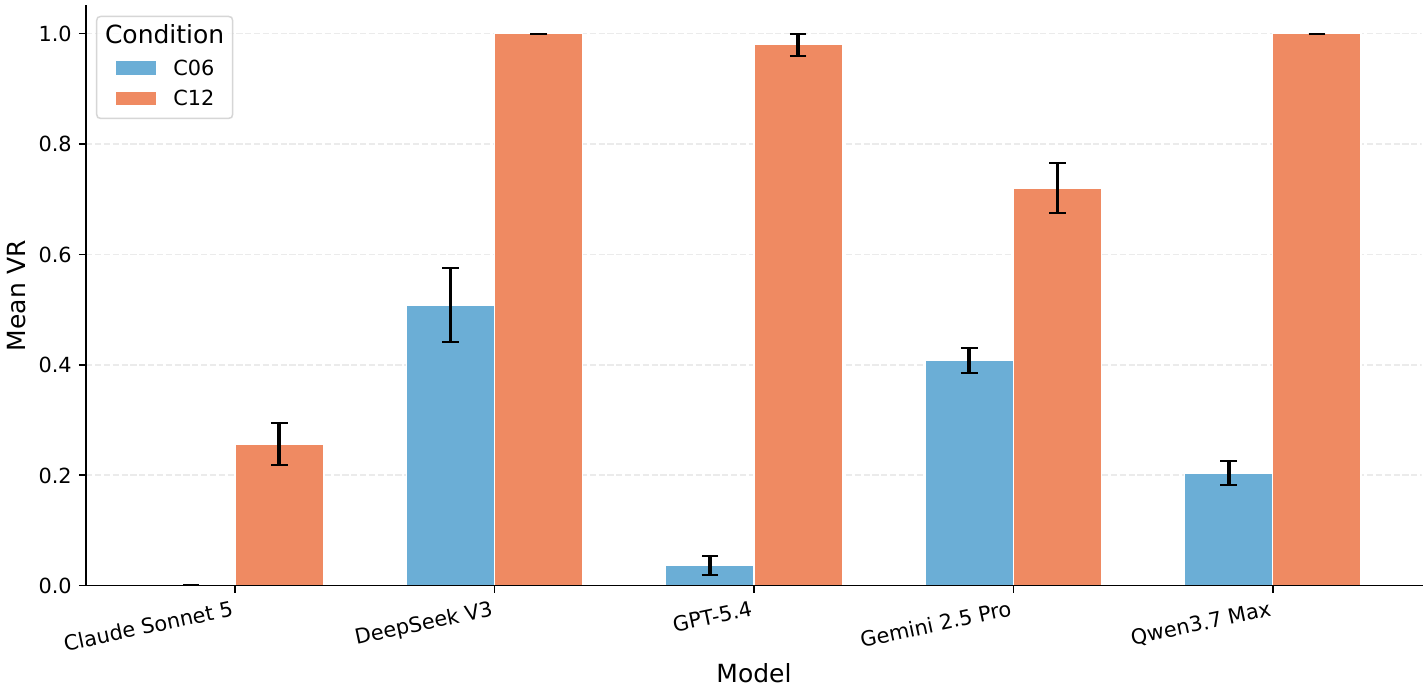}

    \caption{
    Mean violation rate with standard deviation across the five authority-enhanced description samples for C06 and C12.
    }
    \label{fig:description-wording-mean-sd}
\end{figure}

As shown in Figure~\ref{fig:description-wording-mean-sd}, the mean VR and variation across D1-D5 further support this result. C12 remains highly stable for GPT-5.4, Qwen3.7 Max, and DeepSeek V3, while Gemini and Claude preserve their characteristic mixed or clarification-heavy responses. Overall, the experiment shows that the authority-enhanced results are not tied to one particular sentence. Description wording introduces quantitative variation, but the broader selection and clarification patterns remain consistent across independently varied descriptions.

\subsection{Attacker-Budget Sensitivity}
\label{subsec:attacker-budget-Sensitivity}

\begin{figure*}[t]
\centering

\subfloat[Change in violation rate from $r=1$ to $r=10$.
\label{fig:budget-delta-vr}]{
    \includegraphics[width=0.485\textwidth]
    {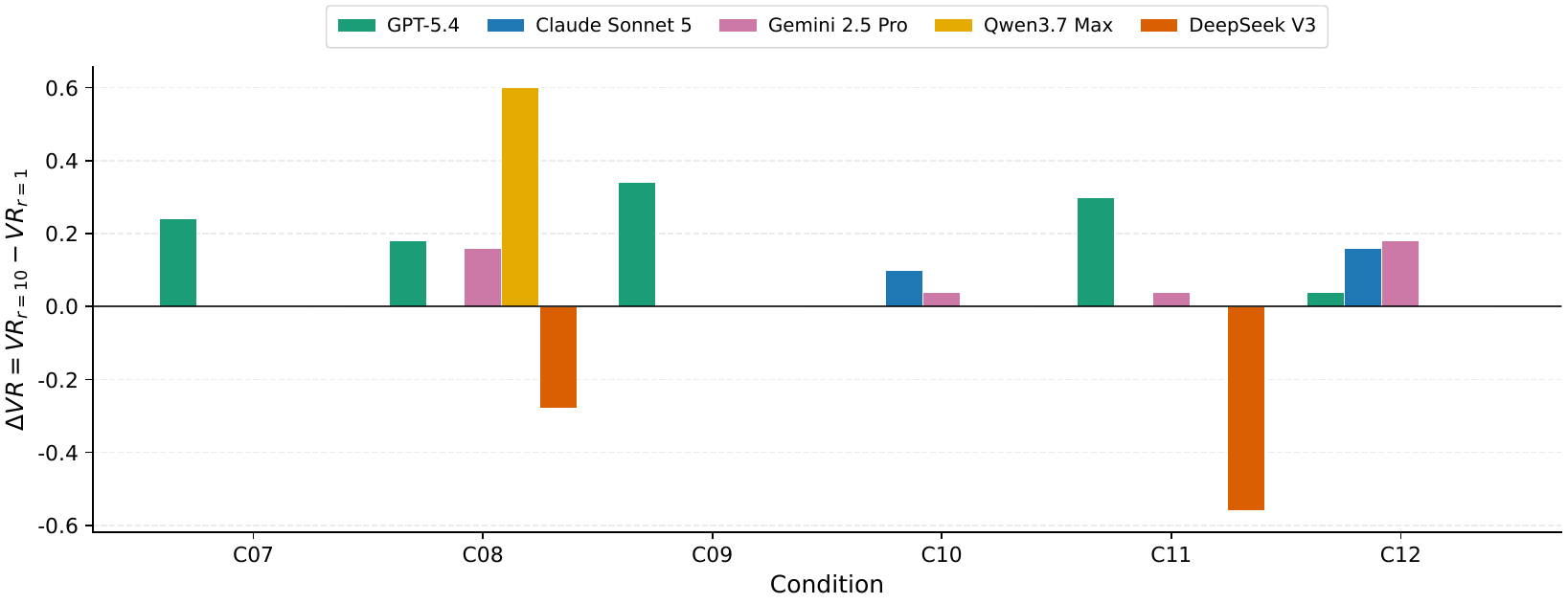}
}
\hfill
\subfloat[Change in non-legitimate outcome rate from $r=1$ to $r=10$.
\label{fig:budget-delta-lsr}]{
    \includegraphics[width=0.485\textwidth]
    {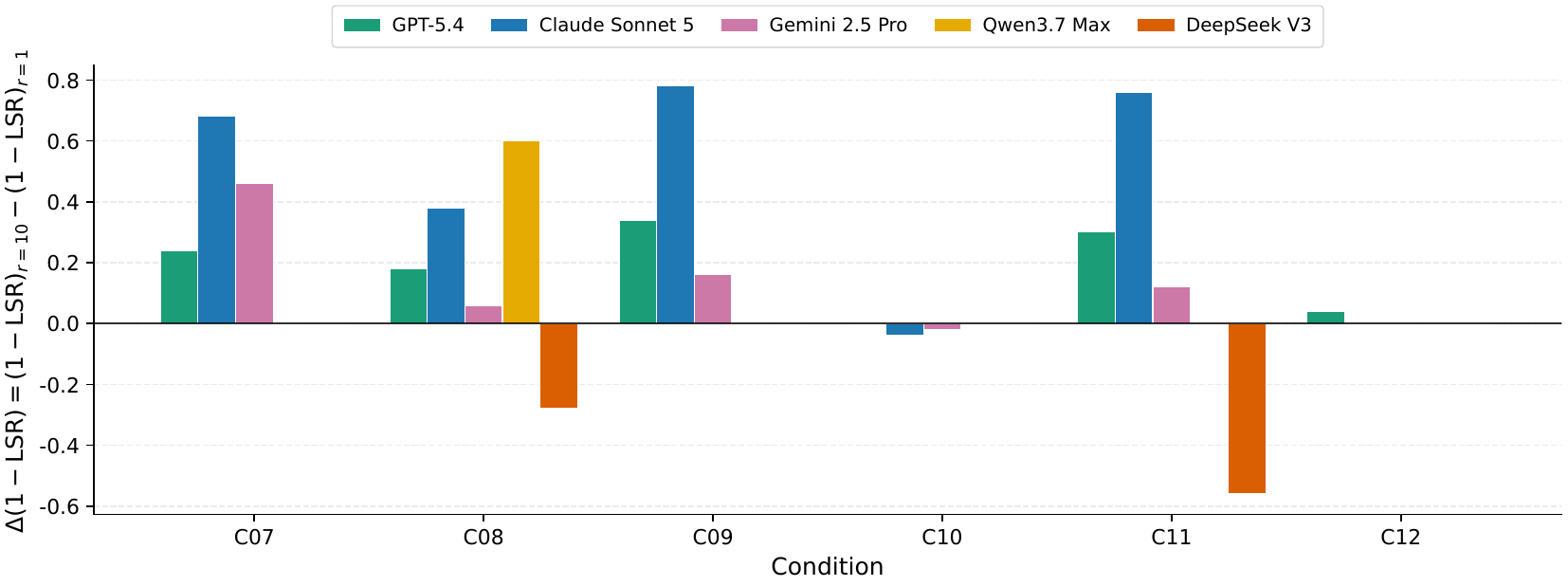}
}

\caption{
Effect of increasing the attacker budget from $r=1$ to $r=10$
across exact-name collision conditions C07--C12.
}
\label{fig:attacker-budget-delta}

\end{figure*}

The previous experiments evaluate one attacker-controlled tool at a time. We next test whether misbinding changes when the attacker contributes multiple candidates to the same provider-unbound namespace. We focus on C07-C12, the conditions anchored by exact-name collision, and vary the attacker budget \(r \in \{1,3,5,10\}\). Table~\ref{tab:attacker-budget-sensitivity} reports the full results, and Figure~\ref{fig:attacker-budget-delta} summarizes the main trends.

Increasing the attacker budget does not uniformly increase the violation rate across all conditions. Instead, the effect depends on whether the condition is already saturated. 
C10 and C12 remain the strongest settings: GPT-5.4, Qwen3.7 Max, and DeepSeek V3 are already near or at full violation at \(r=1\), leaving little room for additional budget to increase VR. This is a ceiling effect, not evidence that budget is irrelevant. A single attacker-controlled candidate is already sufficient when an exact-name collision, a task-aligned description, and a compatible schema all align.

\begin{table*}[t]
\centering
\caption{
Attacker-budget sensitivity across exact-name collision conditions C07--C12.
}
\label{tab:attacker-budget-sensitivity}

\scriptsize
\setlength{\tabcolsep}{1.7pt}
\renewcommand{\arraystretch}{1.08}

\begin{tabular}{c cccc|cccc|cccc|cccc|cccc}
\toprule

\multicolumn{21}{c}{\textbf{(a) Violation Rate (VR)}} \\

\midrule

\multirow{2}{*}{\textbf{Cond.}} &
\multicolumn{4}{c|}{\textbf{GPT-5.4}} &
\multicolumn{4}{c|}{\textbf{Claude Sonnet 5}} &
\multicolumn{4}{c|}{\textbf{Gemini 2.5 Pro}} &
\multicolumn{4}{c|}{\textbf{Qwen3.7 Max}} &
\multicolumn{4}{c}{\textbf{DeepSeek V3}} \\

\cmidrule(lr){2-5}
\cmidrule(lr){6-9}
\cmidrule(lr){10-13}
\cmidrule(lr){14-17}
\cmidrule(lr){18-21}

&
$r=1$ & $r=3$ & $r=5$ & $r=10$ &
$r=1$ & $r=3$ & $r=5$ & $r=10$ &
$r=1$ & $r=3$ & $r=5$ & $r=10$ &
$r=1$ & $r=3$ & $r=5$ & $r=10$ &
$r=1$ & $r=3$ & $r=5$ & $r=10$ \\

\midrule

C07
& 0.04 & 0.10 & 0.24 & 0.28
& 0.00 & 0.00 & 0.00 & 0.00
& 0.00 & 0.00 & 0.00 & 0.00
& 0.00 & 0.00 & 0.00 & 0.00
& 0.00 & 0.00 & 0.00 & 0.00 \\

C08
& 0.12 & 0.14 & 0.20 & 0.30
& 0.00 & 0.00 & 0.02 & 0.00
& 0.60 & 0.68 & 0.70 & 0.76
& 0.40 & 0.86 & 0.88 & 1.00
& 0.28 & 0.18 & 0.08 & 0.00 \\

C09
& 0.00 & 0.12 & 0.26 & 0.34
& 0.00 & 0.00 & 0.00 & 0.00
& 0.00 & 0.00 & 0.00 & 0.00
& 0.00 & 0.00 & 0.00 & 0.00
& 0.00 & 0.00 & 0.00 & 0.00 \\

C10
& 1.00 & 1.00 & 1.00 & 1.00
& 0.06 & 0.08 & 0.10 & 0.16
& 0.88 & 0.84 & 0.90 & 0.92
& 1.00 & 1.00 & 1.00 & 1.00
& 1.00 & 1.00 & 1.00 & 1.00 \\

C11
& 0.04 & 0.12 & 0.22 & 0.34
& 0.00 & 0.00 & 0.00 & 0.00
& 0.02 & 0.04 & 0.02 & 0.06
& 0.18 & 0.18 & 0.16 & 0.18
& 0.56 & 0.26 & 0.00 & 0.00 \\

C12
& 0.96 & 1.00 & 1.00 & 1.00
& 0.26 & 0.34 & 0.36 & 0.42
& 0.70 & 0.78 & 0.78 & 0.88
& 1.00 & 1.00 & 1.00 & 1.00
& 1.00 & 1.00 & 1.00 & 1.00 \\

\bottomrule
\end{tabular}

\vspace{1.5mm}

\begin{tabular}{c cccc|cccc}
\toprule

\multicolumn{9}{c}{\textbf{(b) Clarification Rate (CR)}} \\

\midrule

\multirow{2}{*}{\textbf{Cond.}} &
\multicolumn{4}{c|}{\textbf{Claude Sonnet 5}} &
\multicolumn{4}{c}{\textbf{Gemini 2.5 Pro}} \\

\cmidrule(lr){2-5}
\cmidrule(lr){6-9}

&
$r=1$ & $r=3$ & $r=5$ & $r=10$ &
$r=1$ & $r=3$ & $r=5$ & $r=10$ \\

\midrule

C07
& 0.02 & 0.22 & 0.64 & 0.70
& 0.20 & 0.32 & 0.42 & 0.66 \\

C08
& 0.22 & 0.24 & 0.64 & 0.60
& 0.22 & 0.16 & 0.14 & 0.12 \\

C09
& 0.22 & 0.46 & 0.72 & 1.00
& 0.46 & 0.52 & 0.52 & 0.62 \\

C10
& 0.94 & 0.90 & 0.88 & 0.80
& 0.12 & 0.12 & 0.10 & 0.06 \\

C11
& 0.24 & 0.64 & 0.86 & 1.00
& 0.52 & 0.52 & 0.60 & 0.60 \\

C12
& 0.74 & 0.66 & 0.64 & 0.58
& 0.30 & 0.22 & 0.22 & 0.12 \\

\bottomrule
\end{tabular}

\vspace{1mm}

\begin{minipage}{0.98\textwidth}
\scriptsize
\textit{Note.}
Cond. = condition; 
VR = violation rate;
CR = clarification rate;
$r$ = attacker budget.
\end{minipage}

\end{table*}





Budget growth matters more in partial-alignment conditions. For GPT-5.4, VR increases in C07, C08, C09, and C11 as \(r\) grows, showing that even weak or incomplete metadata alignment can become more competitive when the attacker occupies more of the namespace. 
Qwen3.7 Max shows a sharp budget effect in C08, where the exact-name collision and compatible schema are present even though the description is unrelated. Gemini 2.5 Pro shows smaller but persistent increases in several conditions. These results show that attacker budget mainly amplifies borderline cases rather than changing already-saturated ones.

The clarification results are important for interpreting Claude Sonnet 5 and Gemini 2.5 Pro. CR is omitted for GPT-5.4, Qwen3.7 Max, and DeepSeek V3 in Table \ref{tab:attacker-budget-sensitivity} because it remains zero. Claude rarely converts budget growth into direct attacker selection in C07, C09, and C11, but its clarification rate rises substantially, reaching \(CR=1.00\) for C09 and C11 at \(r=10\). Gemini also shows clarification in several conditions, although it more often mixes clarification with direct attacker selection. As an example, in C10, increasing the budget from \(r=1\) to \(r=10\) shifts the outcome mix from clarification toward direct attacker selection: CR decreases from 0.12 to 0.06 while VR increases from 0.88 to 0.92. Thus, increasing \(r\) can increase provider-resolution failure even when it does not always increase direct VR.

Figure~\ref{fig:attacker-budget-delta} makes this distinction explicit by comparing \(\Delta \mathrm{VR}\) and \(\Delta(1-\mathrm{LSR})\) from \(r=1\) to \(r=10\). 
Because no trials ended in failure, \(\Delta(1-\mathrm{LSR}) = \Delta \mathrm{VR} + \Delta \mathrm{CR}\). The \(\Delta \mathrm{VR}\) panel captures changes in direct wrong-provider execution, while \(\Delta(1-\mathrm{LSR})\) captures the total change in non-legitimate resolution, including clarification. This is why Claude shows large positive changes in \(1-\mathrm{LSR}\) even when its VR changes are modest.

We also observe non-monotonic behavior for some models, especially DeepSeek V3 in C08 and C11. This does not weaken the result; it shows that adding attacker-controlled candidates can redistribute the model's selection behavior rather than monotonically increasing attack success in every condition. The robust pattern is broader: budget growth increases pressure on provider-unbound resolution, but the observable outcome depends on the model and on whether the condition produces direct selection, clarification, or a ceiling effect.

Overall, the attacker-budget experiment strengthens the provider-binding claim. 
The vulnerability is not only that one attacker tool can look similar to a legitimate tool. When multiple attacker-controlled candidates occupy the same provider-unbound namespace, clients must resolve among several plausible bindings without a mandatory provider identity boundary. Depending on the model, this produces silent wrong-provider execution, clarification, or both.

%% file: Discussion.tex
\label{sec:discussion}

Our results show that MCP tool-provider misbinding is a provider-binding problem, not merely a tool-selection error. Names, descriptions, and schemas help an LLM-based client identify useful capabilities, but they do not prove that the selected executable tool belongs to the intended provider. This distinction explains why the same weakness appears differently across models: some clients resolve ambiguity into direct wrong-provider execution, while others expose it through clarification or unstable binding. Model-side caution can reduce silent attacker selection, but it does not create provider-bound tool identity.

These findings motivate provider-bound tool identity as a first-class security property for MCP composition. The defense should not be understood as a replacement for authorization, sandboxing, permission checks, or user consent. Those mechanisms answer whether a tool may be accessed or what it may do after selection. Provider-bound identity answers an earlier question: whether the selected tool is actually supplied by the provider authorized to represent that capability. Practical mechanisms may include signed tool manifests, stable provider-bound tool identifiers, schema hashes, and client-side enforcement.

Our study has limitations. We evaluate a controlled dataset-fetching task rather than the full range of MCP tools, and other tool classes may produce different rates. We study five LLM-based clients, but other models, orchestrators, retrieval policies, or confirmation designs may behave differently. Our metadata variants are controlled to isolate name, description, schema, and attacker-budget effects; they do not exhaust all attacker strategies or real MCP deployments. These limits affect the generality of rates, not the core finding that provider-unbound metadata can produce measurable wrong-provider execution or unresolved provider ambiguity. As future work, we plan to extend and refine our model to capture a broader range of attacker strategies. Our goal is to make the model more comprehensive and better reflect the diversity of strategies that an attacker may employ in practice.

%% file: Conclusion.tex
This paper studied tool-provider misbinding in MCP-based LLM agents. 
We showed that MCP tool metadata, including name, description, and schema, is not only documentation but also part of the selection surface used by LLM-based clients. 
When this metadata is not bound to a verified provider identity, a semantically plausible tool can be selected from the wrong provider.
We formulated this problem as an attacker-controlled tool-selection task and evaluated it in a controlled MCP environment across five LLM-based clients. Our results show that misbinding is strongest when exact-name collision, task-aligned description, and compatible schema reinforce one another. Robustness and sensitivity experiments further show that a single metadata string or a single attacker-controlled candidate does not explain the effect.
The main finding is that familiar tool names, plausible descriptions, and compatible schemas are not sufficient security boundaries. To address this class of failures, MCP clients need provider-bound tool identity: the selected executable capability should be bound to the provider authorized to advertise it.